\documentclass{aa}
\usepackage[varg]{txfonts}
\usepackage{amsmath}
\allowdisplaybreaks[4]
\usepackage{amssymb}
\usepackage{graphicx}
\usepackage{float}
\usepackage{subfigure}
\usepackage{leftidx}
\usepackage{CJK}
\usepackage{esint}
\usepackage{bm}
\usepackage{color}
\usepackage{tabu}
\usepackage{booktabs}
\usepackage{threeparttable}
\usepackage{multirow}
\usepackage{longtable}
\usepackage{rotating}
\usepackage{makecell}

\usepackage{url}
\usepackage[colorlinks, linkcolor=blue, anchorcolor=blue, citecolor=blue, filecolor=blue, menucolor=blue, runcolor=blue, urlcolor=blue]{hyperref}
\begin{document}
  
  \title{Modelling carbon-chain species formation in lukewarm corinos with new multi-phase models}
  %\subtitle{}
  \author{Yao Wang\inst{1,2}\and Qiang Chang\inst{3}\and
Hongchi Wang\inst{1}}
  \institute{Purple Mountain Observatory and Key Laboratory of Radio Astronomy, Chinese Academy of Sciences, 8 Yuanhua Road, Nanjing 210034, China; \email{wangyao@pmo.ac.cn}
  \and
University of Science and Technology of China, 96 Jinzhai Road, Hefei 230026, China
  \and
Xinjiang Astronomical Observatory, Chinese Academy of Sciences, 150 Science 1-Street, Urumqi 830011, China; \email{changqiang@xao.ac.cn}
  }
  \date{Received date month year / Accepted date month year}
  
  %\begin{document}
  %\maketitle
  
    \abstract{Abundant carbon-chain species have been observed towards lukewarm corinos L1527, B228, and L483. These carbon-chain species are believed to be synthesized in the gas phase after CH$_4$ desorbs from the dust grain surface at the temperature around 30 K. }{We investigate carbon-chain species formation in lukewarm corinos using a more rigorous numerical method and advanced surface chemical models. We also pay attention to the significance of the finite size effect. }{We used the macroscopic Monte Carlo method in our simulations. In addition to the two-phase model, the basic multi-phase model and the new multi-phase models were used for modelling surface chemistry on dust grains. All volatile species can sublime at their sublimation temperatures in the two-phase model while most volatile species are frozen in the ice mantle before water ice sublimes in the basic and the new multi-phase models. The new multi-phase models allow more volatile species to sublime at their sublimation temperatures than the basic multi-phase model does. }{The significance of the finite size effect is dependent on the duration of the cold phase. The discrepancies between the rate equation approach and the Monte Carlo method decrease as the duration of the cold phase increases. When $T \sim 30\, \mathrm{K}$, the abundances of gaseous CH$_4$ and CO in the two-phase model are the highest while the basic multi-phase model predicts the lowest CO and CH$_4$ abundances among all models. The abundances of carbon-chain species in the basic and the new multi-phase models are lower than that in the two-phase model when $T \sim 30\, \mathrm{K}$ because CH$_4$ is crucial for the synthesis of carbon-chain species. However, because the abundance of electrons increases as the abundance of H$_3$O$^+$ decreases, some carbon-chain species abundances predicted by the basic multi-phase model may not be lower than that in the new multi-phase models. The two-phase model performs best in predicting carbon-chain species abundances to fit observations while the basic multi-phase model works the worst. The abundances of carbon-chain species predicted by the new multi-phase models agree reasonably well with observations. }{The amount of CH$_4$ that can diffuse inside the ice mantle, thus sublime upon warm-up plays a crucial role in the synthesis of carbon-chain species in the gas phase. The carbon-chain species observed in lukewarm corinos may be able to gauge surface chemical models. }

   
   %\titlerunning{?????}
   \maketitle

  \section{Introduction} \label{section1}
  
  \par{}Abundant carbon-chain molecules have been observed in the early stage of low-mass star forming regions which are known as lukewarm corinos. The mechanism for synthesizing carbon-chain molecules in these sources is known as warm carbon-chain chemistry (WCCC) \citep{2008ApJ...672..371S}, therefore the lukewarm corinos with abundant carbon-chain molecules are also called WCCC sources. L1527 in Taurus is confirmed to be the first WCCC source \citep{2008ApJ...672..371S}, which contains the low-mass protostar IRAS 04368+2557 in the transitional phase from Class 0 to Class I \citep{2000prpl.conf...59A, 2008ApJ...681.1385H, 2009ApJ...702.1025S, 2010ApJ...722.1633S}. So far 14 carbon-chain species with relatively high abundances have been observed towards L1527. IRAS 15398-3359 in Lupus, which is also identified as B228, is confirmed as the second WCCC source \citep{2009ApJ...697..769S}. This protostar is also in the transitional phase from Class 0 to Class I. Six carbon-chain species have been observed towards B228. Recently, abundant carbon-chain species as well as more saturated complex organic molecules (COMs) are observed towards L483 \citep{2017ApJ...837..174O} in the Aquila Rift. This cloud contains the Class 0 protostar IRAS 18148-0440 \citep{1995ApJ...453..754F}. Other possible WCCC sources include B1, L1448N \citep{2009ApJ...697..769S, 2011ApJ...743..182H}, and the L1251A molecular core around the Class 0 protostar L1251A IRS3 \citep{2011ApJ...730L..18C}. 

  \par{}The discovery of abundant carbon-chain species in low-mass star forming regions is unexpected because these species are characteristics of dark clouds \citep{1992ApJ...392..551S, 2008ApJ...681.1385H, 2008ApJ...672..371S}. Before abundant carbon-chain species were observed towards L1527, COMs have been observed towards low-mass star forming regions \citep{2003ApJ...593L..51C, 2004ApJ...617L..69B, 2005ApJ...632..973J, 2006PASJ...58L..15S, 2007A&A...463..601B, 2012ApJ...757L...4J, 2014A&A...563L...2M, 2015ApJ...804...81T}. These low-mass star forming regions with COMs are known as hot corinos, which are warmer than the lukewarm corinos. The desorption of CH$_4$ is the first step to synthesize carbon-chain species in the lukewarm corinos according to the WCCC mechanism \citep{2008ApJ...672..371S, 2009ApJ...697..769S, 2016ApJ...833L..14L}. When the temperature arises to about 30 K, volatile species such as CH$_4$ can sublime from the grain surface. The sublimated species participate in gas phase reactions to produce carbon-chain species. So far, carbon-chain species abundances predicted by chemical models for L1527 agree reasonably well with observations \citep{2008ApJ...674..984A, 2008ApJ...681.1385H, 2011ApJ...743..182H}. However, further modelling studies on WCCC are necessary because of the limitations of previous models. 
  \par{}First, some previous models use the rate equation (RE) approach in the numerical simulations which in some cases may be problematic. When the average population of reactive species on the grain surface is well below one, the RE approach can lead to large errors in the simulation results \citep{1997OLEB...27...23T}. This problem is known as the finite size effect. \citet{2009ApJ...691.1459V} showed that the abundances of species predicted by the RE approach deviate much from that by the more rigorous macroscopic Monte Carlo (MC) approach if the temperature is within 20 K and 50 K. Although previous WCCC model results at about 30 K agree with observations \citep{2008ApJ...681.1385H}, it is questionable whether the model results simulated with the more rigorous numerical approach still fit the observations well. 

  \par{}Second, a more serious problem with previous WCCC simulations is the way their models treat the surface chemistry. All volatile species such as CH$_4$ and CO in the ice mantle of a dust grain are assumed to sublime when the temperature is higher than their sublimation temperatures in these models. However, laboratory experiments show that not all CO can desorb when the temperature is higher than the CO sublimation temperature \citep{1988Icar...76..201S, 2003ApJ...583.1058C}. There is also observational evidence that CO might be trapped in the ice mantle. For instance, the abundance of CO in some prestellar cores is observed to be lower than its canonical value \citep{2012A&A...540A..75F}, and one possible explanation is that CO is trapped in the ice mantle. As the sublimation of CH$_4$ triggers the synthesis of carbon-chain species in the gas phase, the production of carbon-chain species may be less efficient if a fraction of CH$_4$ molecules are locked in the ice mantle. 
  
  \par{}More advanced surface chemical models are necessary so that the trapped CH$_4$ in the mantle at temperatures higher than 30 K is considered. All previous WCCC models have adopted the widely used two-phase model in which all surface species can participate in reactions including sublimation. In order to consider the trapping effect, a more advanced model called the three-phase model was suggested by \citet{1993MNRAS.263..589H}. This model distinguishes species on the grain surface active layers and species buried in the ice mantle. Species in the active layers are reactive while species in the ice mantle are chemically inert. As the temperature increases, only species in the active layers can sublime, thus the majority of solid species are trapped in the ice mantle. The original three-phase model was simulated using the RE approach, but recently, \citet{2013ApJ...762...86V} performed more rigorous macroscopic MC simulations of the three-phase model. However, the assumption in the three-phase model that species in the bulk of ice are completely inert may not be correct. For example, it is estimated that the probability that photons are absorbed by one monolayer of ice is about 0.007 \citep{2008A&A...491..907A}. Thus photons can easily penetrate through about 150 monolayers of ice and photo-dissociate species deeply buried in the ice mantle. A few modified three-phase models that adopt a partially active bulk ice mantle are suggested \citep{2013ApJ...765...60G, 2014ApJ...787..135C, 2017MNRAS.467.1763K, 2018arXiv180902419L}. Moreover, in the model suggested by \citet{2018arXiv180902419L}, the exponential decay with depth of photolysis in the ice mantle is well mimicked. 

  \par{}In this work, we have investigated WCCC using the macroscopic MC method and the advanced surface chemical models. We are particularly interested in studying the carbon-chain species abundances predicted by different surface models. The models used in this work are the two-phase model, the three-phase model suggested by \citet{2013ApJ...762...86V}, and the new multi-phase models suggested by \citet{2018arXiv180902419L}. 

  \par{}In Sect. \ref{section2} we describe our chemical models and the physical model of lukewarm corinos. In Sect. \ref{section3} we present all model results. In Sect. \ref{section4} we compare the results with observations. In Sect. \ref{section5} we discuss the advantages and disadvantages of different models and present our conclusions.

  \section{Model} \label{section2}
      \subsection{Chemical model} \label{section2.1}

        \begin{table*}
        \caption{Model parameters. }
        \label{table1}
        \begin{center}
          \begin{tabular}{lll}\\
          \hline
          \hline
            Parameter&Symbol&Value\\
            \hline
            Sticking coefficient&$P_{\mathrm{s}}$&0.5\\
            Reactive desorption efficiency&$a_{\mathrm{RRK}}$&0.01\\
            Ratio of surface diffusion barriers to desorption energies&$E_{\mathrm{b1}}/E_{\mathrm{d}}$&0.5\\
            Ratio of bulk diffusion barriers to desorption energies&$E_{\mathrm{b2}}/E_{\mathrm{d}}$&0.7\\
            Barrier thickness&$b$&1 \r{A}\\
            Grain radius&$r$&0.1 $\mu$m\\
            Dust grain density&$\rho_{\mathrm{d}}$&3 g cm$^{-3}$\\
            Dust-to-gas mass ratio&$m_{\mathrm{d/g}}$&0.01\\
            Number of absorption sites on each grain&$S$&1.0$\times$10$^6$\\
            Surface density of sites&$N_{\mathrm{S}}$&7.96$\times$10$^{14}$ sites cm$^{-2}$\\
            Unattenuated FUV flux&$\chi$&$\chi_0=1$\\
            Cosmic ray ionization rate&$\zeta_{\mathrm{CR}}$&1.3$\times$10$^{-17}$ s$^{-1}$\\
            Fraction of the timescale between two cosmic-ray heating events&$f$&3.16$\times$10$^{-19}$\\
            Characteristic vibrational frequency&$\nu_0$&1$\times$10$^{12}$ s$^{-1}$\\
          \hline\\
          \end{tabular}\\
        \end{center}
        \end{table*}
        
        \begin{table}
        \caption{Initial abundances \citep{2008ApJ...681.1385H}. }
        \label{table2}
        \begin{center}
          \begin{tabular}{ll}\\
          \hline
          \hline
            Species&$n_{\mathrm{X}}/n_{\mathrm{H}}$\tablefootmark{a}\\
            \hline
            H$_{2}$&5.00(-1)\\
            He&1.40(-1)\\
            C$^{+}$&7.30(-5)\\
            N&2.14(-5)\\
            O&1.76(-4)\\
            S$^{+}$&8.00(-8)\\
            Na$^{+}$&2.00(-9)\\
            Mg$^{+}$&7.00(-9)\\
            Si$^{+}$&8.00(-9)\\
            P$^{+}$&3.00(-9)\\
            Cl$^{+}$&4.00(-9)\\
            Fe$^{+}$&3.00(-9)\\
          \hline\\
          \end{tabular}\\
          \tablefoot{\tablefoottext{a}{$a(b)=a\times10^b .$}}
        \end{center}
        \end{table}

      \par{}The new multi-phase model used in this work has been explained in detail by \citet{2018arXiv180902419L}. Here we only briefly explain the model in the following. The ice mantle on the grain surface is composed of surface active layers and a partially active bulk of ice. Species in the surface active layers can diffuse and participate in surface reactions including sublimation. There are two types of binding sites in the partially active bulk of ice, the normal sites and the interstitial sites. Species occupying normal sites are called normal species, while interstitial species occupy interstitial sites and are uniformly distributed in the partially active bulk of ice. Normal species are locked in the mantle, thus cannot sublime until the whole ice mantle sublimes, while interstitial species can diffuse in the bulk of ice and sublime at their sublimation temperatures. Interstitial species can be generated by the photolysis of species in the bulk of ice. Both the UV photons from the external radiation and that induced by cosmic ray are considered in our chemical model. Photo-fragments of icy species can enter normal sites or interstitial sites. The probability for a photo-dissociation product at the M-th layer of ice to enter the interstitial site is, 
\begin{align}
P = \frac{\alpha N_{emp, int}}{\alpha N_{emp, int} +N_{emp, nor, M}},  \label{equation1}
\end{align}
where $N_{emp, int}$ is the average number of empty interstitial sites, $N_{emp, nor, M}$ is the number of empty normal sites at the M-th layer, and $\alpha$ is a parameter depending on the difference of the binding energy of normal sites and interstitial sites. The larger the parameter $\alpha$ is, the more likely that photo-dissociation products occupy interstitial sites and, therefore, the more interstitial volatile species can sublime as the temperature increases. However, the parameter $\alpha$ is poorly known so far. \citet{2018arXiv180902419L} argued that $\alpha$ should be no more than one, thus we set $\alpha$ to be 1, 0.5, and 0.01 in the new multi-phase models. Moreover, we also simulate a new multi-phase model in which all photo-dissociation products enter interstitial sites ($\alpha =+\infty$) to maximize the population of interstitial species in the model. New multi-phase models adopting $\alpha =+\infty$, 1, 0.5, and 0.01 are called new multi-phase model 0, 1, 2, and 3 in this work, respectively. We also simulated the two-phase model and the three-phase model suggested by \citet{2013ApJ...762...86V}, which is referred to as the basic multi-phase model in this work. The amounts of CH$_4$ that can sublime at the CH$_4$ sublimation temperature are different in these models. In the basic multi-phase model, CH$_4$ molecules only in the active layers can sublime while all surface CH$_4$ can desorb in the two-phase model. The amount of CH$_4$ molecules that can sublime in the new multi-phase model is larger than that in the basic multi-phase model but less than that in the two-phase model, because CH$_4$ molecules in the active layers and those occupying interstitial sites can sublime at the CH$_4$ sublimation temperature in the new multi-phase models. Therefore, among all the models, carbon-chain species in the two-phase model should be the most abundant while the abundances of carbon-chain species predicted by the basic multi-phase model should be the lowest. 
      \par{}The macroscopic MC method is used to simulate the two-phase model, the basic multi-phase model and the new multi-phase models. Following \citet{2009ApJ...691.1459V}, we isolate a cell of gas surrounding a dust particle. The total population of H nuclei is $10^{12}$ in the cell of gas if we set $\rho_{\mathrm{d}}$ = 3 g cm$^{-3}$ and $m_{\mathrm{d/g}}$ = 0.01 \citep{2009ApJ...691.1459V}. In this work, we report the fractional abundance of species, which is defined as the ratio of the population of the species to total population of H nuclei (10$^{12}$) in the cell. If the population of a species is one, its fractional abundance is 10$^{-12}$. This abundance is the minimum fractional abundance resolvable if we run a model only once. In order to reduce statistical fluctuation, we ran each model ten times to calculate the average fractional abundances, so the lowest fractional abundance resolvable is 10$^{-13}$. Any fractional abundance lower than 10$^{-13}$ is reported to be zero in this paper. In order to research the finite size effect, we also use the RE approach to simulate the two-phase model for comparison. 
      \par{}We performed the macroscopic MC simulations of the gas-grain chemical reaction network used by \citet{2008ApJ...681.1385H} (hereafter HHG), which is based on the OSU2008 reaction network but with some changes, and found that even in the two-phase model, carbon-chain species may be severely underproduced. The abundances of carbon-chain species in the two-phase model using a reaction network that includes all reactions in the OSU2008 and HHG agree with observations better than that using OSU2008 or HHG individually, therefore the combined reaction network is utilized in all models in this work. As a result, the gas-grain chemical reaction network includes 6462 reactions and 655 species. The competition mechanism is utilized to calculate the reaction rates for two surface reactions with barriers, CO + H $\rightarrow$ HCO and H$_2$CO + H $\rightarrow$ H$_3$CO \citep{2012ApJ...759..147C}. The reaction barriers of these two reactions are taken from \citet{2009A&A...505..629F}. Finally, the accretion of H$_2$ consumes too much CPU time in the macroscopic MC simulations. However, it is found that H$_2$ accretion is not important for the molecular evolution of dense clouds \citep{2017ApJ...851...68C}. Thus, H$_2$ accretion is not included in our chemical reaction network in order to reduce the computational cost. The bulk diffusion barrier for an interstitial species is set to be $E_{\mathrm{b2}} = 0.7E_{\mathrm{d}}$ \citep{2018arXiv180902419L}, where $E_{\mathrm{d}}$ is the desorption energy of the same species in the active layers. All other parameters in the network are the same as that used by \citet{2008ApJ...681.1385H} and \citet{2010A&A...522A..42S}, which are shown in Table \ref{table1}. We adopted the low initial metal abundances as shown in Table \ref{table2} \citep{2008ApJ...681.1385H}.

        \begin{table*}
          \caption{$\Delta t$ and the corresponding temperatures when the abundances of HC$_5$N, HC$_7$N and HC$_9$N reach $10^{-12}$ in the two-phase model adopting different $t_{\mathrm{cold}}$ values. }
          \label{table3}
          \begin{center}
          \begin{tabular}{ccccccc}\\
          \hline
          \hline
            \multirow{2}{*}{\makecell{$t_{\mathrm{cold}}$ (yr)\tablefootmark{a}}} &
            \multicolumn{2}{c}{HC$_5$N} &
            \multicolumn{2}{c}{HC$_7$N} &
            \multicolumn{2}{c}{HC$_9$N}  \\
            \cline{2-3}
            \cline{4-5}
            \cline{6-7} 
            %&$t$&$T$& \multicolumn{2}{c}{} 
            %$t$&$T$& \multicolumn{2}{c}{}
            %$t$&$T$& \multicolumn{2}{c}{} \\
            &$\Delta t$ (yr)\tablefootmark{b}&$T$ (K)\tablefootmark{c}&$\Delta t$ (yr)&$T$ (K)&$\Delta t$ (yr)&$T$ (K) \\
            \hline
            MC $1.0\times 10^5$&$6.97\times 10^4$&33.08&$7.72\times 10^4$&38.31&$9.96\times 10^4$&57.12 \\
            MC $2.0\times 10^5$&$6.01\times 10^4$&27.16&$7.05\times 10^4$&33.61&$7.55\times 10^4$&37.08 \\
            MC $3.0\times 10^5$&$4.78\times 10^4$&20.85&$5.60\times 10^4$&24.90&$5.81\times 10^4$&26.03 \\
            MC $4.0\times 10^5$&$4.26\times 10^4$&18.62&$5.60\times 10^4$&24.90&$5.72\times 10^4$&25.54 \\
            MC $5.0\times 10^5$&$4.55\times 10^4$&19.83&$5.67\times 10^4$&25.27&$5.97\times 10^4$&26.93 \\
            MC $6.0\times 10^5$&$4.50\times 10^4$&19.62&$5.83\times 10^4$&26.14&$5.93\times 10^4$&26.70 \\
            MC $7.0\times 10^5$&$4.53\times 10^4$&19.75&$5.79\times 10^4$&25.92&$5.96\times 10^4$&26.87 \\
            MC $8.0\times 10^5$&$4.46\times 10^4$&19.45&$5.82\times 10^4$&26.09&$6.16\times 10^4$&28.02 \\
            MC $9.0\times 10^5$&$5.07\times 10^4$&22.21&$5.87\times 10^4$&26.37&$6.29\times 10^4$&28.79 \\
            MC $1.0\times 10^6$&$4.57\times 10^4$&19.92&$5.75\times 10^4$&25.70&$6.10\times 10^4$&27.67 \\
          \hline
            RE $1.0\times 10^5$&$5.40\times 10^4$&23.85&$5.68\times 10^4$&25.32&$5.96\times 10^4$&26.87 \\
            RE $3.0\times 10^5$&$5.23\times 10^4$&22.99&$5.87\times 10^4$&26.37&$5.87\times 10^4$&26.37 \\
          \hline
          \end{tabular}
        \end{center}
        \tablefoot{
          \tablefoottext{a}{$t_{\mathrm{cold}}$ represents the timescale of the cold phase. MC represents the Monte Carlo simulation and RE represents the rate equation calculation. }
          \tablefoottext{b}{$\Delta t$ represents the timescale from the beginning of the warm-up phase to the time when the abundance of HC$_5$N, HC$_7$N, and HC$_9$N begins to be over $10^{-12}$. }
          \tablefoottext{c}{$T$ is the temperature of the core at the time $t_{\mathrm{cold}}$ + $\Delta t$. }
        }
        \end{table*}

      \subsection{Physical model} \label{section2.2}
      \par{}We adopted the physical model used by \citet{2008ApJ...681.1385H}, which is a homogeneous one-point model that includes two stages, a cold phase when the temperature of the molecular cloud is 10 K at the beginning and then a warm-up phase. In this model, the total hydrogen nucleus density is $n_{\mathrm{H}}=10^6 \, \mathrm{cm}^{-3}$ in the central region of the prestellar core and the visual extinction $A_{\mathrm{V}}=10$. The initial temperature is $T=10 \, \mathrm{K}$ meanwhile the gas temperature and the grain temperature are in equilibrium all the time for simplicity. The timescale of the cold phase $t_{\mathrm{cold}}$, when the temperature of the molecular cloud remains to be 10 K, is not well constrained. \citet{2008ApJ...681.1385H} argued through rate equation simulations that their two-phase model results are robust for the range $10^4$ yr $<t_{\mathrm{cold}}<$ $10^6$ yr, thus they suggest a moderate value, $t_{\mathrm{cold}} = 10^5 \, \mathrm{yr}$. However, we find that the results of the two-phase model simulated with the macroscopic MC method are dependent on $t_{\mathrm{cold}}$. Therefore, we simulate the two-phase model with the macroscopic MC method with different $t_{\mathrm{cold}}$ and select a moderate value of $t_{\mathrm{cold}}$ so that the predicted carbon-chain species abundances agree reasonably well with observations. Following \citet{2006A&A...457..927G} and \citet{2008ApJ...681.1385H, 2011ApJ...743..182H}, the temperature at the time $t_{\mathrm{cold}}$ + $\Delta t$ in the warm-up phase is, 
        \begin{align}
        T=T_0+(T_{\mathrm{max}}-T_0) \left( \frac{\Delta t}{t_{\mathrm{h}}} \right) ^n,  \label{equation2}
        \end{align}
        where $T_0$ is the initial temperature, 10 K, $T_{\mathrm{max}}$ is the maximum temperature at the end of the warm-up phase that is adopted to be 200 K for the formation of hot cores or hot corinos, $t_{\mathrm{h}}$ is the heating timescale, and $n$ is the order of the heating. According to \citet{2006A&A...457..927G} and \citet{2008ApJ...681.1385H, 2011ApJ...743..182H}, $t_{\mathrm{h}}=2\times 10^5\, \mathrm{yr}$ as the timescale for intermediate star formation which is derived from the observed protostellar luminosity function \citep{2000A&A...355..617M}, and $n=2$ for power law temperature profiles \citep{2006A&A...457..927G}. After the warm-up phase, the temperature remains constant.

  \section{Results} \label{section3}
  
    \par{}In this section, we report the fractional abundances of carbon-chain species in different models. The finite size effect in the two-phase model is investigated in Sect. \ref{section3.1} through a comparison of the simulation results from the macroscopic MC method and the RE approach. The two-phase, the basic and the new multi-phase models are simulated with the macroscopic MC method, and the results from these models are compared in Sect. \ref{section3.2}.

    \subsection{Finite size effect} \label{section3.1}

        \begin{figure*}
        %\centering
        %\scalebox{1}{ 
        \begin{center}
           %\includegraphics[width=8.8cm]{twophase-all-1-2.pdf}
           %\resizebox{\hsize}{!}{\includegraphics{twophase-all-1-2.pdf}}
           \includegraphics[width=9cm]{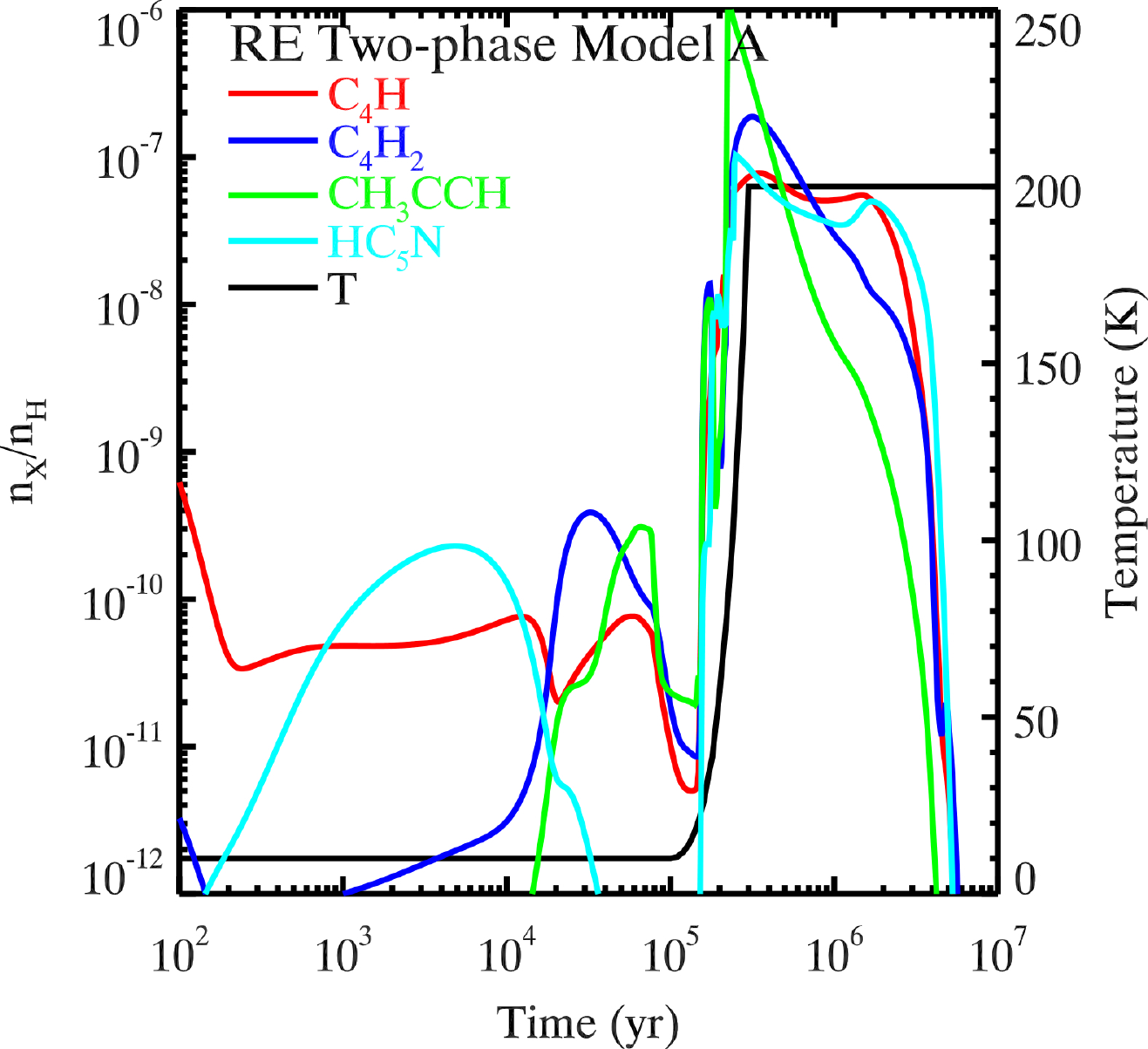}\includegraphics[width=9cm]{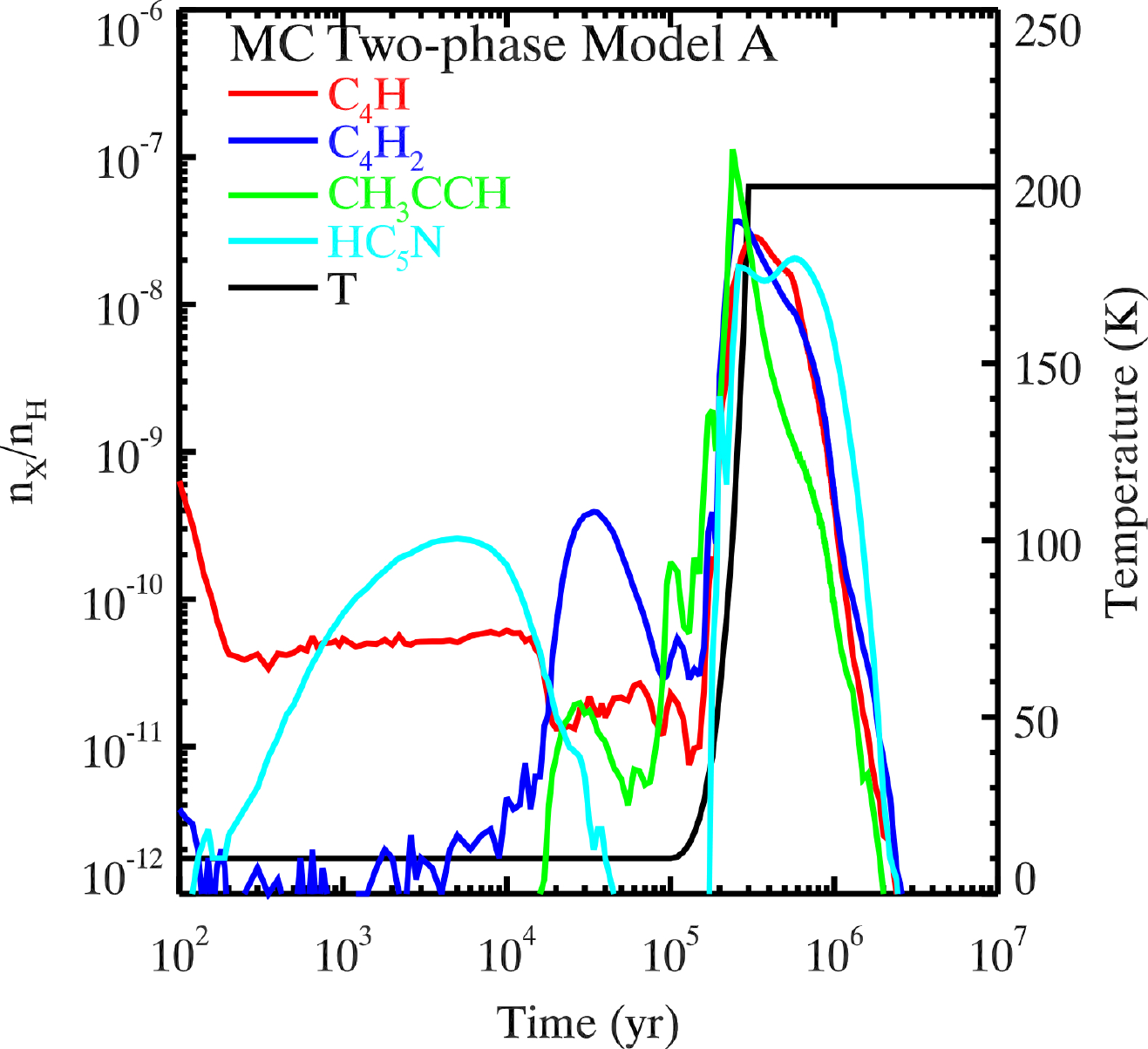} 
           \includegraphics[width=9cm]{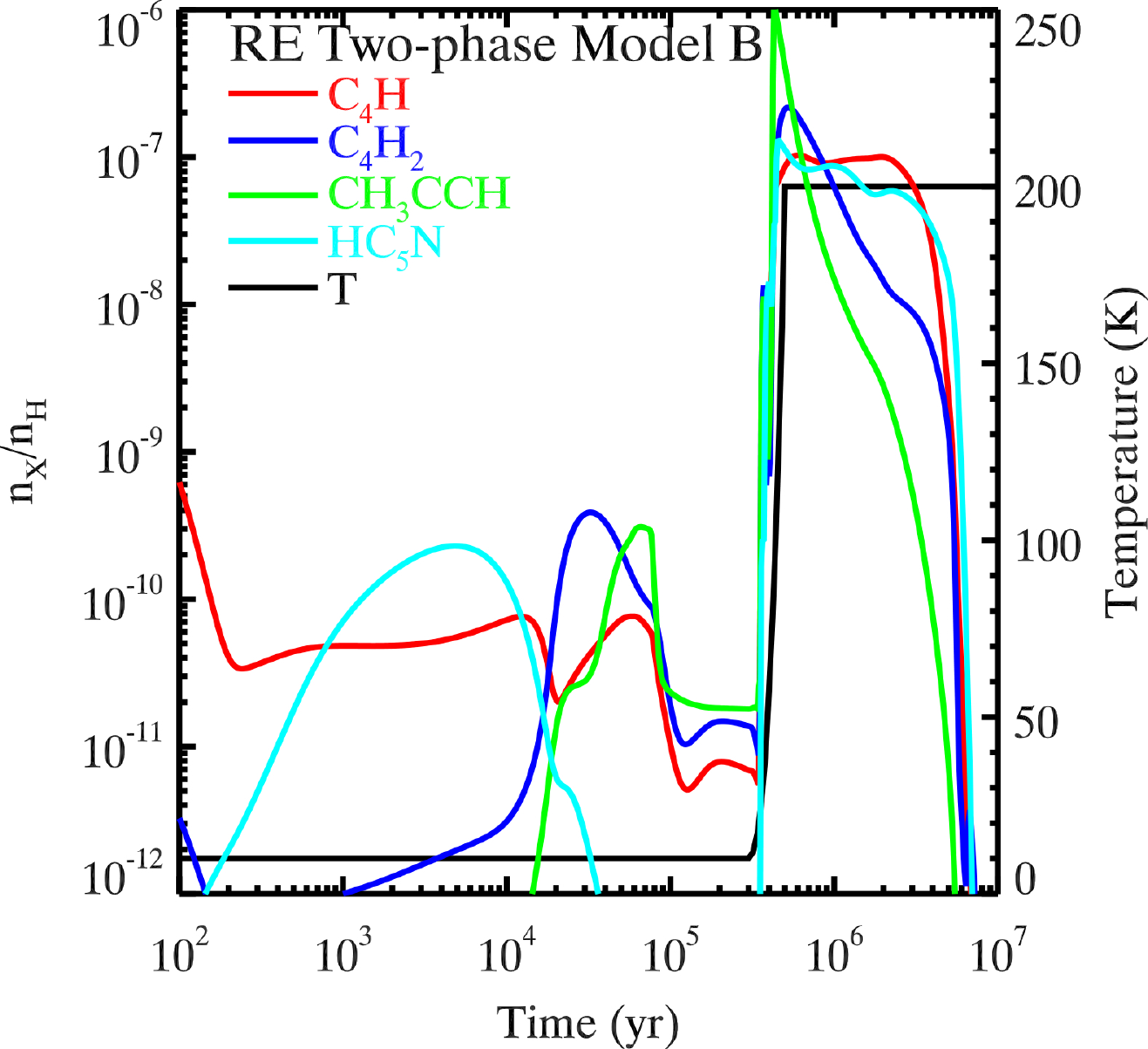}\includegraphics[width=9cm]{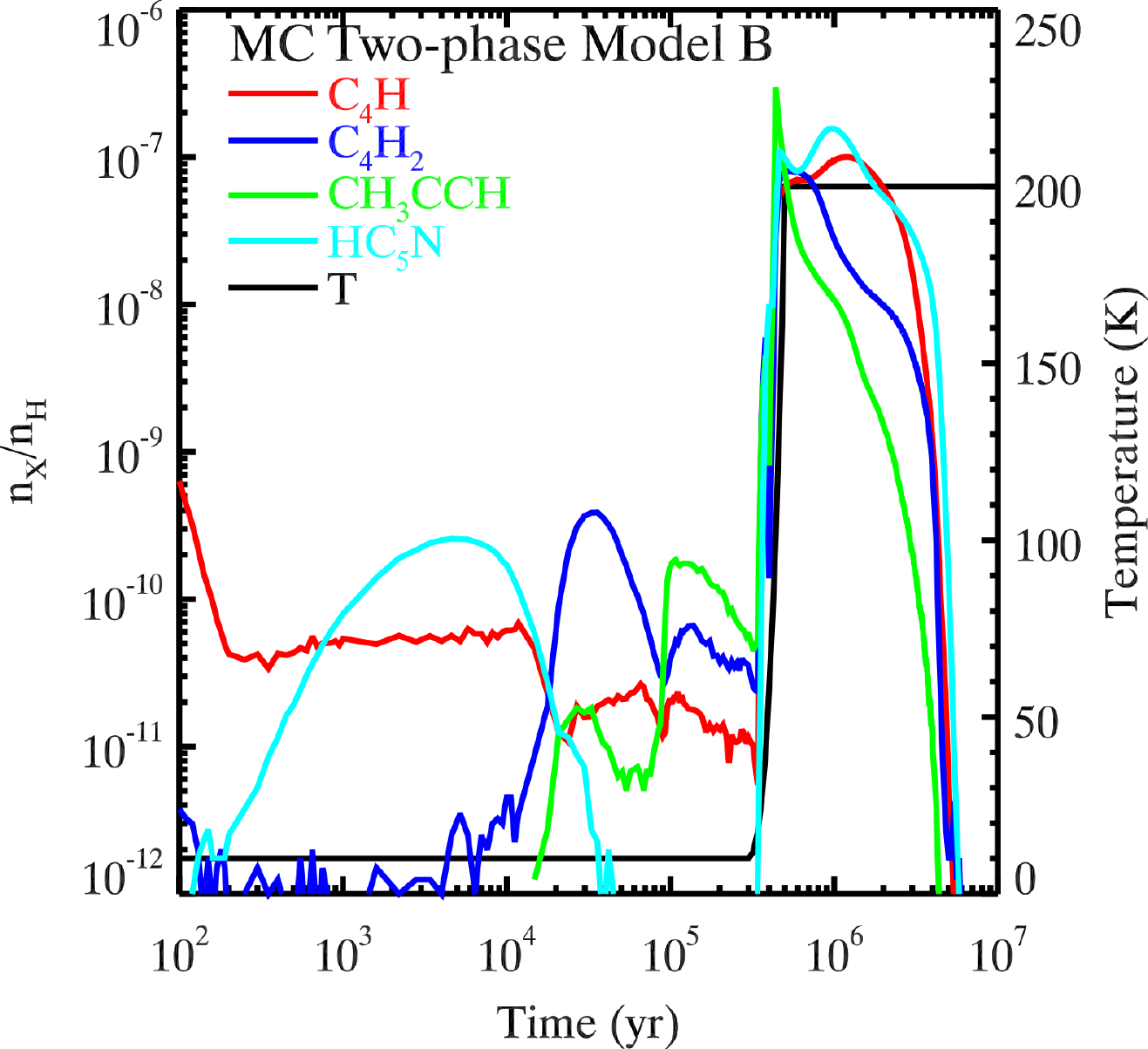} 
        \end{center}
        \caption{Temporal evolution of the fractional abundances of selected carbon-chain species in the two-phase model simulated with the RE approach and the macroscopic MC method. RE represents the rate equation calculation and MC represents the macroscopic Monte Carlo simulation. The two-phase model adopts different $t_{\mathrm{cold}}$, $t_{\mathrm{cold}}=10^5$ yr in the two-phase model A while $t_{\mathrm{cold}}=3\times 10^5$ yr in the two-phase model B. The temporal evolution of the temperature $T$ is also plotted in the figure. }
        \label{figure1}
           %}
        \end{figure*}   
  
  \par{}The two-phase model is simulated with both the macroscopic MC method and the RE approach. The abundances of gas-phase HC$_5$N, HC$_7$N, and HC$_9$N in the two-phase model calculated with the RE approach deviate much from that with the macroscopic MC method. Table \ref{table3} shows $\Delta t$ in the warm-up phase and the corresponding temperatures when the abundance of HC$_5$N, HC$_7$N, and HC$_9$N begins to be higher than $10^{-12}$ adopting different $t_{\mathrm{cold}}$. If $t_{\mathrm{cold}}=10^5$ yr, the temperature has reached 33 K when the abundance of HC$_5$N is over $10^{-12}$ in the simulation with the macroscopic MC method. The temperature can be even higher for HC$_7$N and HC$_9$N, which increases to 38 K and 57 K, respectively. The time at which the fractional abundance is higher than $10^{-12}$, $t_{\mathrm{cold}}$ + $\Delta t$, becomes shorter so that the corresponding temperature is lower if the RE approach is used. For instance, the abundance of HC$_9$N starts to be higher than $10^{-12}$ when the temperature is 27 K if $t_{\mathrm{cold}}=10^5$ yr. These differences arise from the fact that the macroscopic MC method takes into account the finite effect while the RE approach does not \citep{2009ApJ...691.1459V}. Our results agree with previous studies in that the discrepancies of some gas phase carbon-chain species abundances between the macroscopic MC method and the RE approach can be significant when $T=25-30 \, \mathrm{K}$ \citep{2009ApJ...691.1459V}. Therefore, the finite size effect is important if $t_{\mathrm{cold}}=10^5$ yr. On the other hand, the finite size effect becomes less important for longer $t_{\mathrm{cold}}$. If $t_{\mathrm{cold}}=3\times 10^5$ yr, the temperatures of the core are about 21 K, 25 K, and 26 K, respectively, when the abundances of HC$_5$N, HC$_7$N, and HC$_9$N are over $10^{-12}$ in the simulations with the macroscopic MC method. The RE approach predicts similar temperatures of the core, which are 23 K, 26 K, and 26 K, respectively. Moreover, in our simulations the temperatures are relatively robust if $t_{\mathrm{cold}} > 3\times 10^5$ yr. The reason for the decrease of the discrepancy between the RE and the macroscopic MC model results as $t_{\mathrm{cold}}$ increases is that as the population of surface species increases, it becomes less likely that the average number of a surface species drops well below one, thus the finite size effect becomes less significant. In order for the abundances of carbon-chain species to be higher than $10^{-12}$ when the temperatures of the core are around 30 K in the MC models, $t_{\mathrm{cold}}$ should be no shorter than $3\times 10^5$ yr. Therefore, we set $t_{\mathrm{cold}}$ to be $3\times 10^5$ yr for all models in Sect. \ref{section3.2}.

        \begin{figure*}
        %\centering
        %\scalebox{1}{ 
           \begin{center}
           \includegraphics[width=9cm]{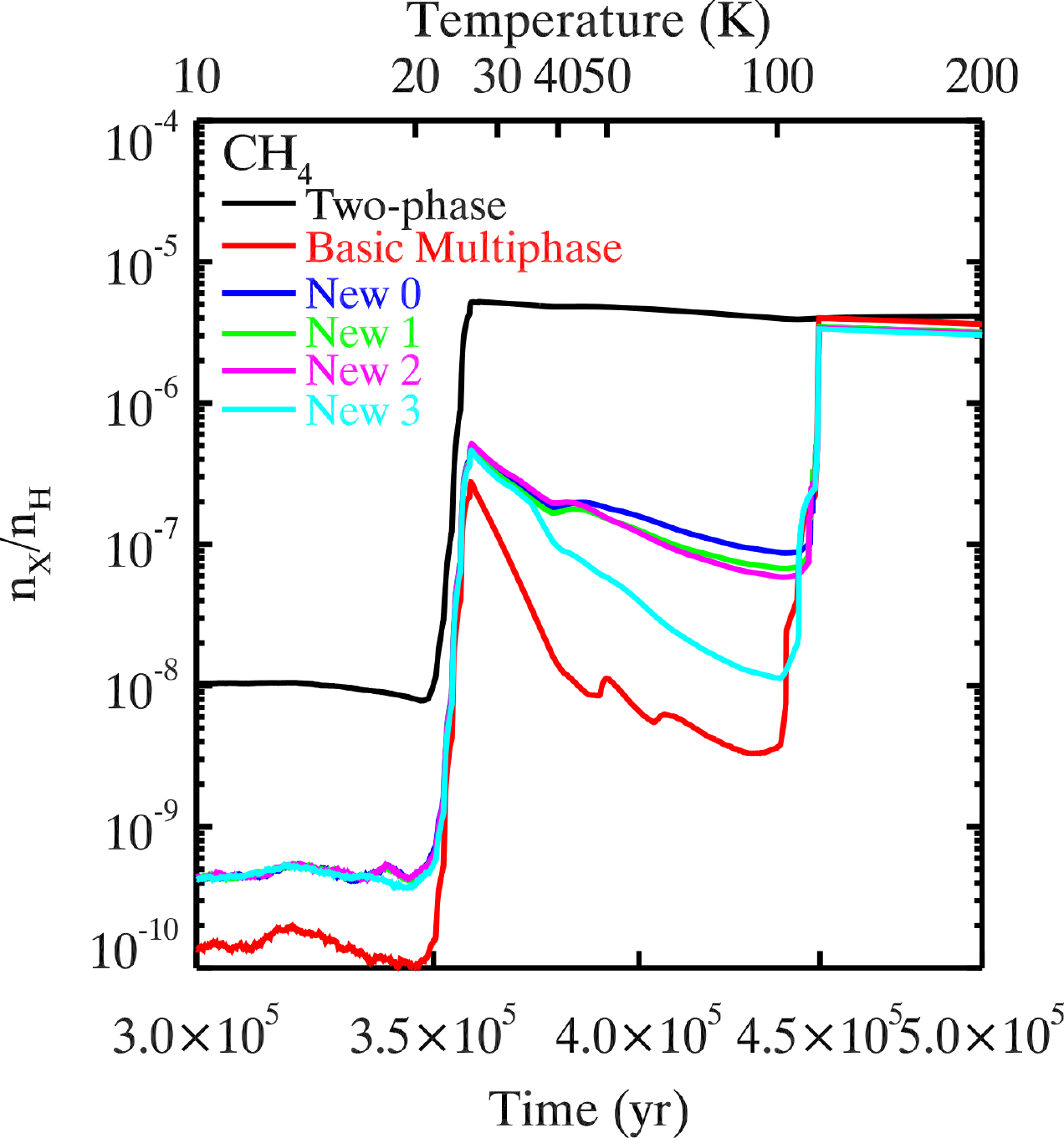}\includegraphics[width=9cm]{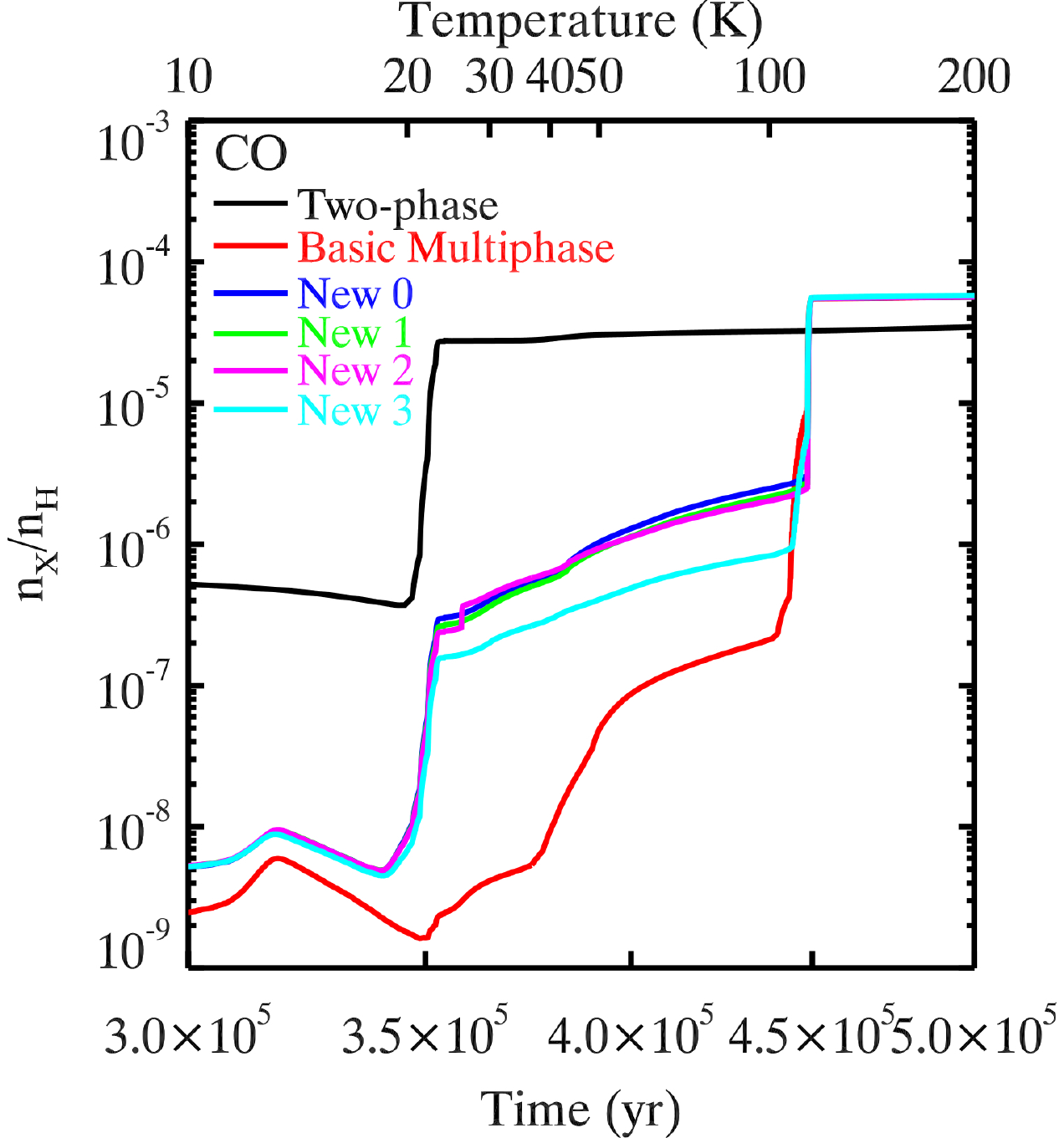} 
           \includegraphics[width=9cm]{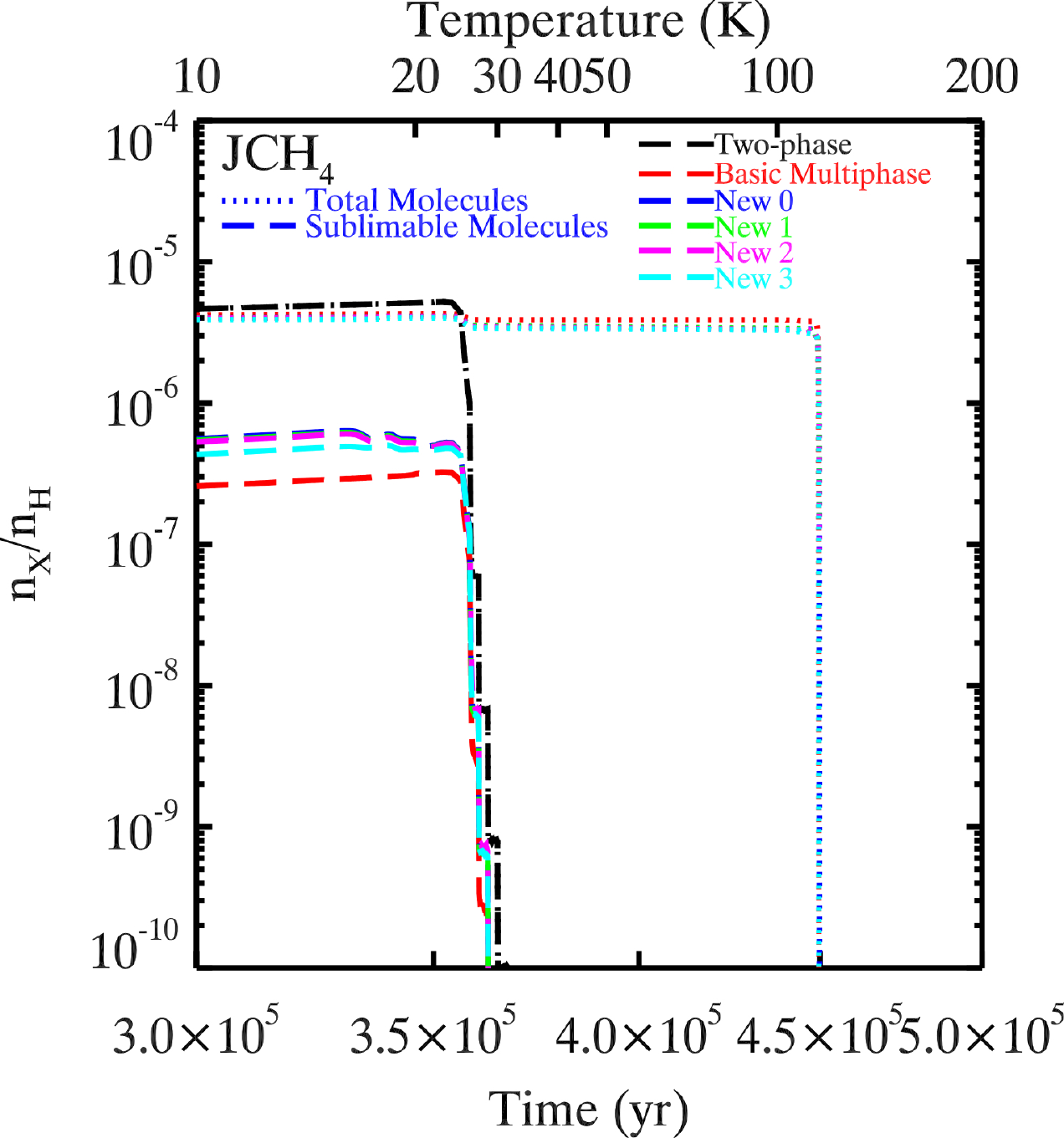}\includegraphics[width=9cm]{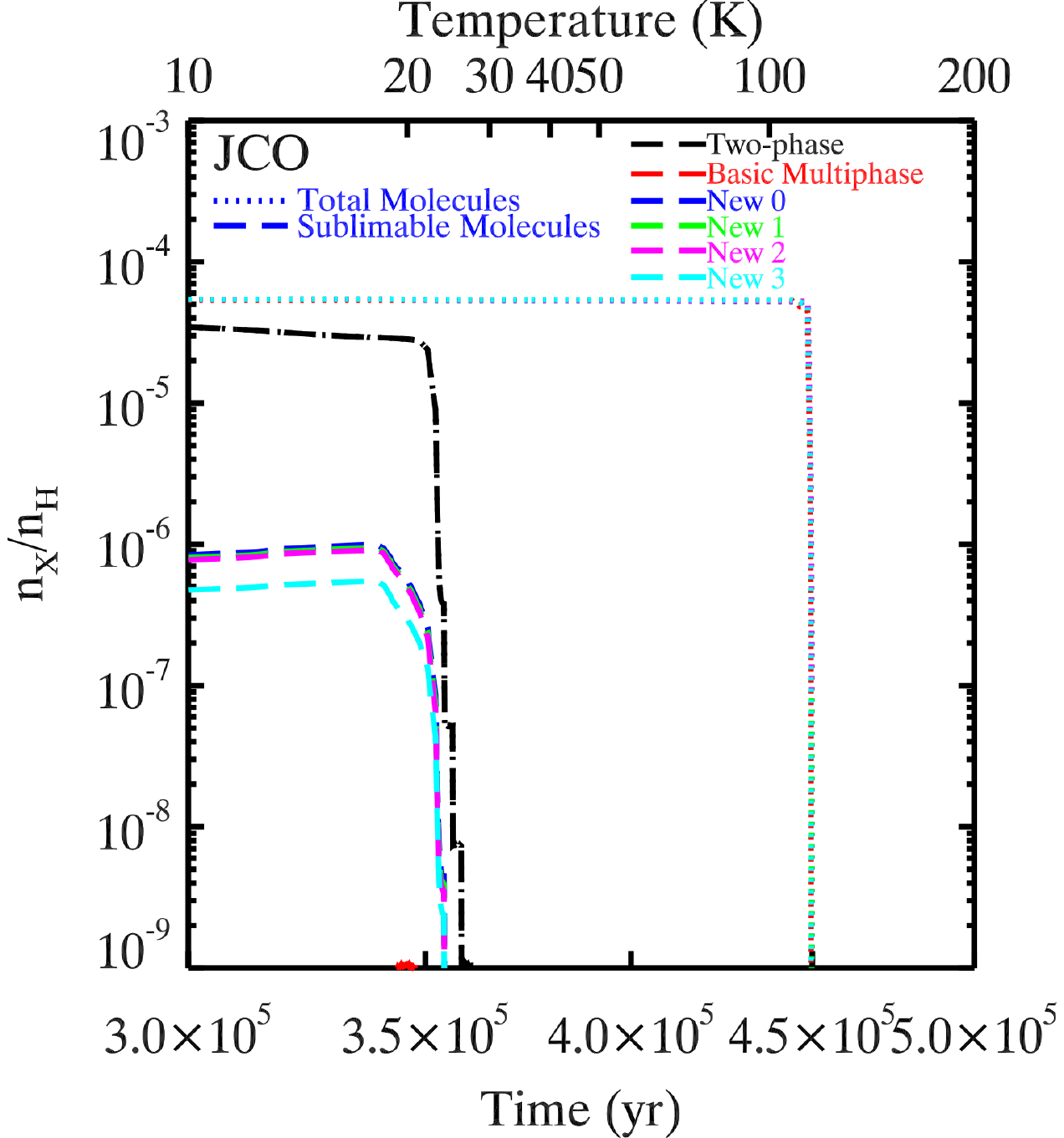}
           \end{center}
           %\caption{}
           %}
           \caption{Temporal evolution of the fractional abundances of CH$_4$, CO, total JCH$_4$, sublimable JCH$_4$, total JCO, and sublimable JCO in all models. The letter J represents species on the grain. The temperature is labelled on the top horizontal axis. For total JCH$_4$ and JCO, the curves from all models except the two-phase model are almost the same. }
           \label{figure2}
        \end{figure*}

        \begin{figure*}
        %\centering
        %\scalebox{1}{ 
        \begin{center}
           \includegraphics[width=9cm]{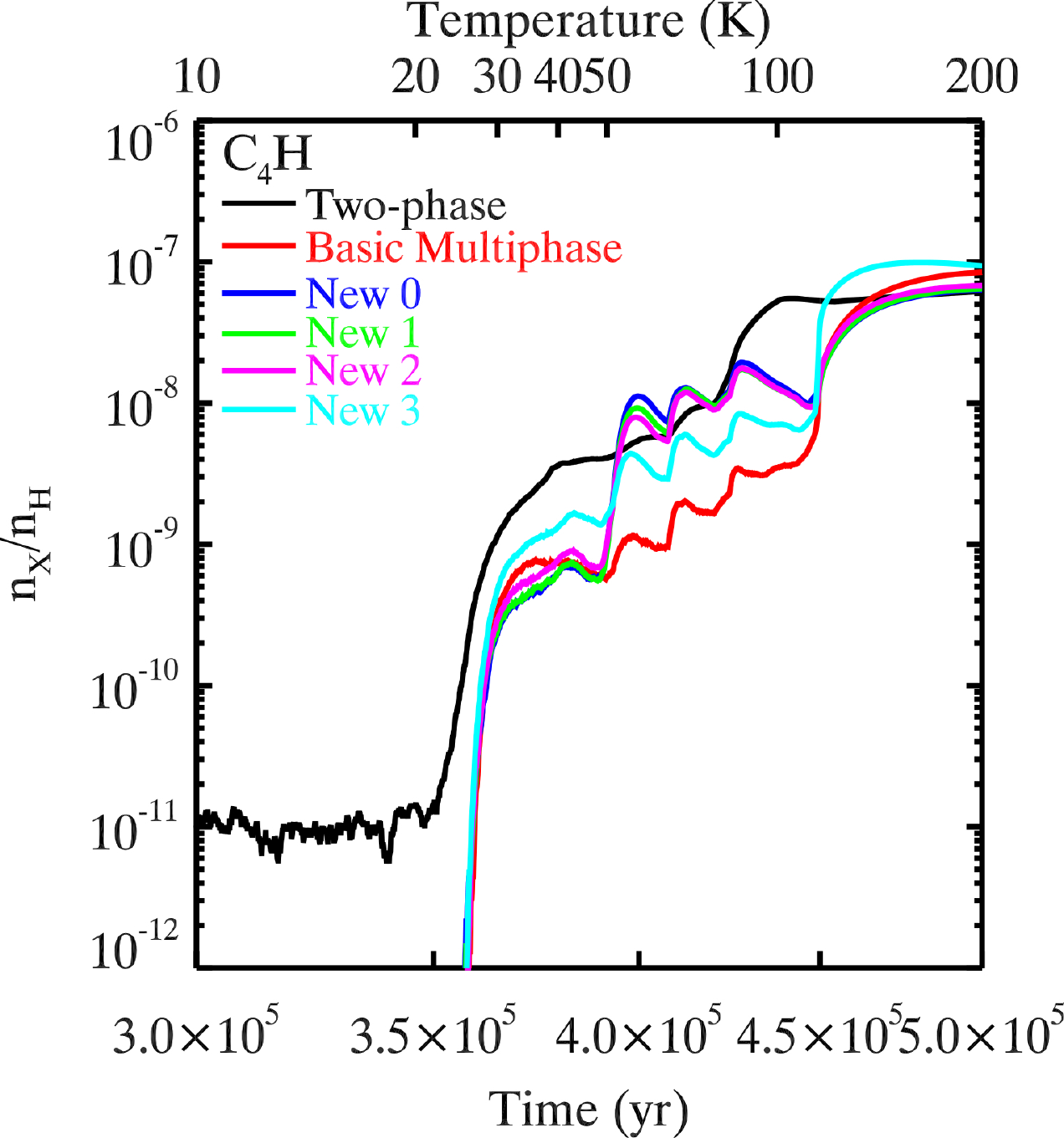}\includegraphics[width=9cm]{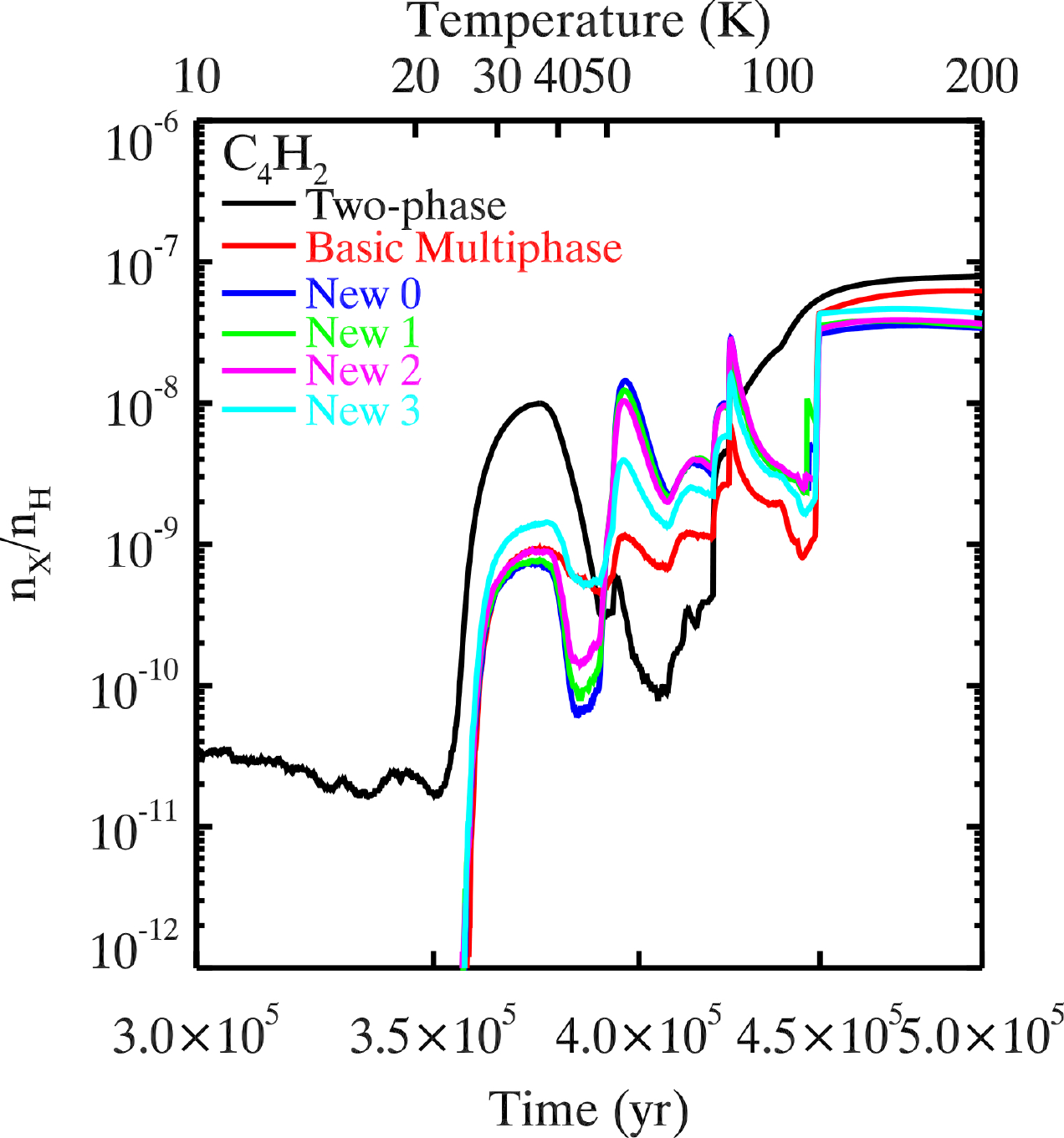} 
           \includegraphics[width=9cm]{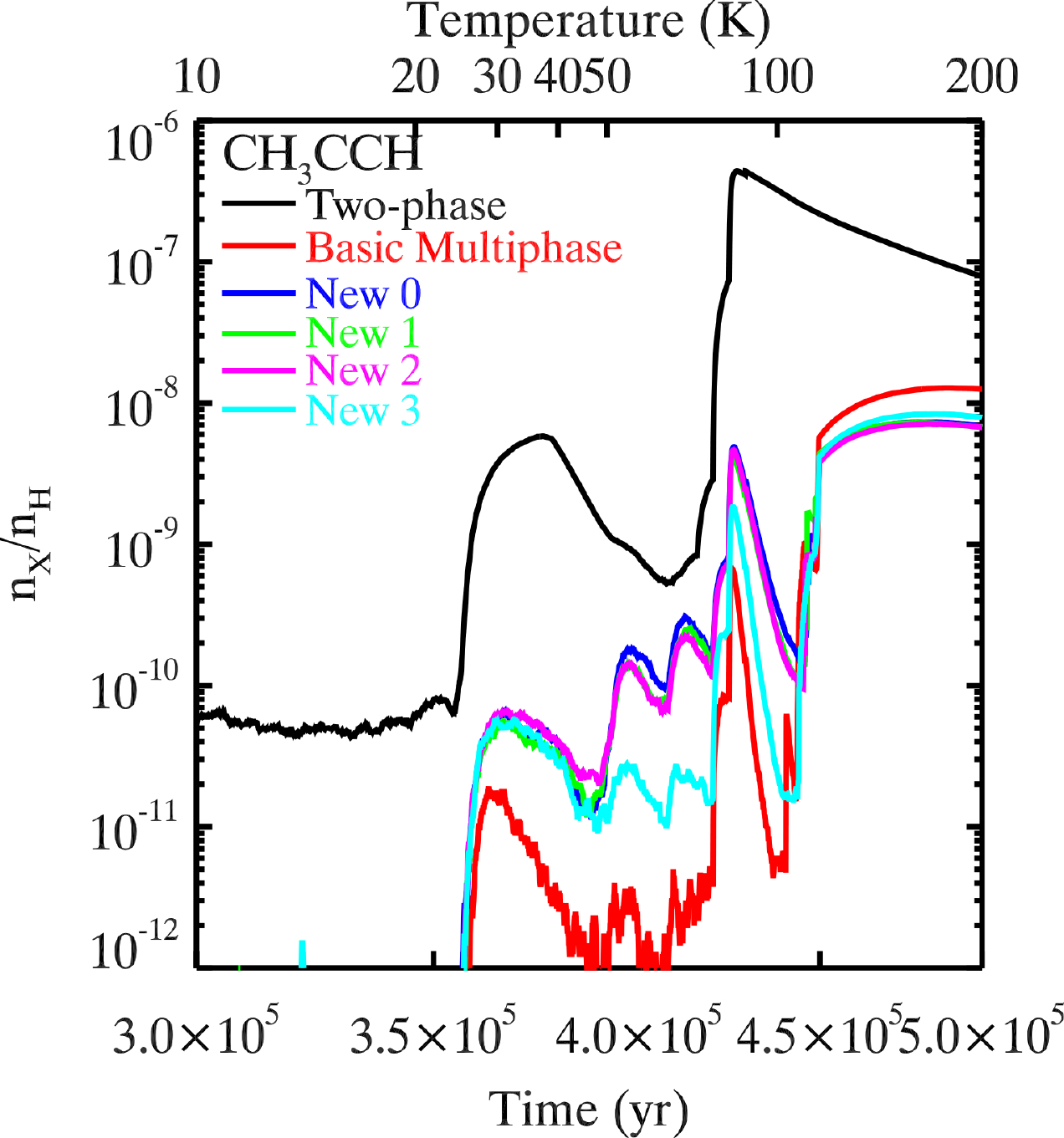}\includegraphics[width=9cm]{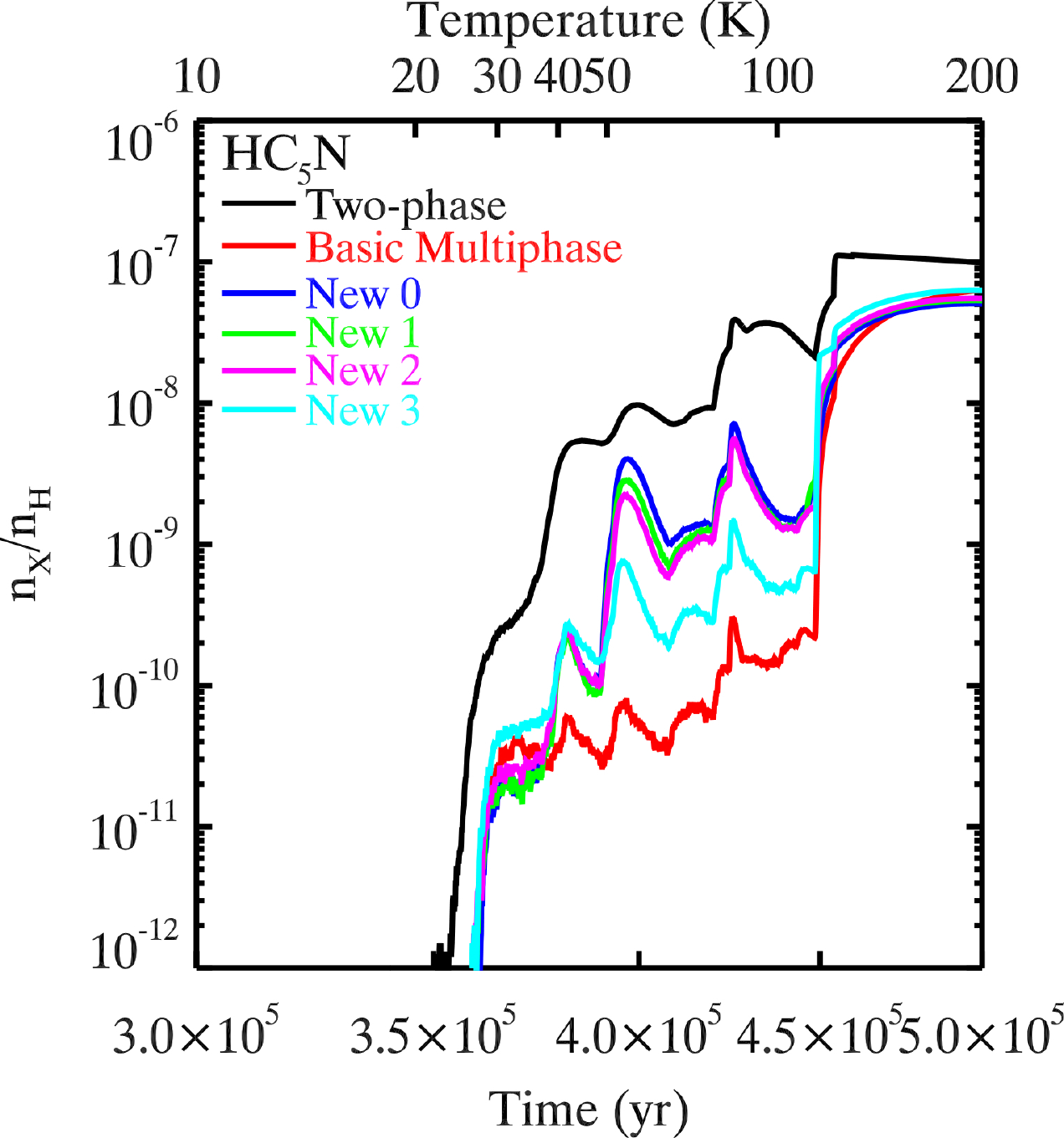} 
           \end{center}
           %\caption{}
           %}
            \caption{Temporal evolution of the fractional abundances of C$_4$H, C$_4$H$_2$, CH$_3$CCH, and HC$_5$N in all models. The temperature is also labelled on the top horizontal axis. }
            \label{figure3}
        \end{figure*}

  \par{}Figure \ref{figure1} shows the evolutions of the abundances of selected gas-phase carbon-chain species in the two-phase model simulated with the RE approach and the macroscopic MC method. Different $t_{\mathrm{cold}}$ is used to perform the simulations. In the two-phase model A, $t_{\mathrm{cold}}$ is set to be $10^5$ yr while in the two-phase model B, it is $3\times 10^5$ yr. In an overview, the abundances of the selected carbon-chain species in the cold stage are much affected, regardless of $t_{\mathrm{cold}}$, by the numerical approach used in simulations. The finite size effect is not important at $T= 10$ K in our models, which agrees with \citet{2009ApJ...691.1459V}. The maximum abundances in the cold stage can reach around $4.0\times 10^{-10}$, which are relatively low in the entire evolution of 10$^7$ yr. On the other hand, the abundances of carbon-chain species rise rapidly in the warm-up phase. The abundances can be more than two orders of magnitude than the peak values in the cold phase. Thus we can conclude that these carbon-chain species are synthesized rapidly during this period. Moreover, in the hot phase with $T=200 \, \mathrm{K}$, the abundances decrease from the peak values to 10$^{-12}$. There are large differences between the results from the RE model and the macroscopic MC model in the warm-up phase when $t_{\mathrm{cold}}=10^5$ yr. The maximum abundance of CH$_3$CCH can be around $10^{-6}$ in the two-phase model A when the RE approach is utilized. However, the maximum abundance of CH$_3$CCH drops by one order of magnitude in the same model when the macroscopic MC method is used. Moreover, it takes a shorter time for the abundances of carbon-chain species to decrease to 10$^{-12}$ when the macroscopic MC method is utilized. In the RE calculations, the abundances of all selected carbon-chain species drop to 10$^{-12}$ at around $t \sim 5 \times 10^6$ yr while in the MC simulations, carbon-chain species drop earlier at around $t \sim 2\times 10^6$ yr. Thus, the finite size is important when $t_{\mathrm{cold}}=10^5$ yr. On the other hand, the discrepancy between the RE model and the macroscopic MC model becomes smaller when longer $t_{\mathrm{cold}}$ is adopted. For instance, the maximum abundance of CH$_3$CCH in the two-phase model B simulated with the RE approach is only about a factor of three larger than that with the macroscopic MC method, compared with an order of magnitude in the case of a shorter $t_{\mathrm{cold}}$.

      \subsection{Model results comparison} \label{section3.2}

        \par{}Hereafter, surface species are designated by the letter J. Surface species that can sublime at their sublimation temperatures before the whole ice mantle disappears are called sublimable species. In the two-phase model all surface species are sublimable species, while in the basic multi-phase model only species in the active layers are sublimable species. In the new multi-phase models, in addition to species in the active layers, interstitial species are also sublimable species. According to \citet{2008ApJ...672..371S, 2009ApJ...697..769S}, CH$_4$ is the precursor for producing carbon-chain molecules, and CO is the precursor for producing complex organic molecules. Therefore, in order to investigate the synthesis of carbon-chain species in each model, it is important to research the abundances of sublimable JCH$_4$ in each model. Figure \ref{figure2} shows the temporal evolution of CH$_4$, CO, total JCH$_4$, sublimable JCH$_4$, total JCO, and sublimable JCO in all models in the warm-up phase. The abundance of gas-phase CH$_4$ in the two-phase model can reach $5.2\times 10^{-6}$ when $T$ is slightly lower than 30 K, but its abundances in the new multi-phase model 0, 1, 2, and 3 can only reach around $3.0\times 10^{-7}$, which are about one order of magnitude lower than that in the two-phase model. The abundance of gaseous CH$_4$ in the basic multi-phase model is even lower, around $1.0\times 10^{-7}$ when $T$ is 30 K. On the other hand, the abundances of gaseous CH$_4$ in all models are almost the same at 200 K when all species in the ice mantle sublime. The difference of gaseous CH$_4$ abundance at 30 K originates from the different populations of sublimable JCH$_4$ molecules in these models. Although the populations of total JCH$_4$ in all models are approximately the same before these molecules start to sublime, the population of sublimable JCH$_4$ molecules in the two-phase model is about one order of magnitude higher than that in the basic multi-phase model, while in the new multi-phase models, it is a factor of between two and three larger than that in the basic multi-phase model. The populations of sublimable JCH$_4$ in the new multi-phase model 0-2 are approximately the same, while in the new multi-phase model 3 it is the lowest among all new multi-phase models. The same phenomenon occurs for CO. The abundance of gaseous CO in the two-phase model at 25 K is much higher than those in the basic and new multi-phase models, while gaseous CO abundances in the new multi-phase models are higher than that in the basic multi-phase model. Similarly, the abundance of sublimable JCO is the highest in the two-phase model and the lowest in the basic multi-phase model at temperatures below 25 K, and the abundances in the new multi-phase models are moderate. 
        \par{}Figure \ref{figure3} shows the temporal evolutions of the abundances of selected gas-phase carbon-chain species in the two-phase, the basic multi-phase, and the new multi-phase models. We pay most attention to the time period when WCCC occurs at the temperature around 30 K \citep{2008ApJ...681.1385H, 2011ApJ...743..182H}. The abundances of all selected gas-phase carbon-chain species in the new multi-phase models 0-2 are similar, which demonstrates that if $\alpha \ge 0.5$, the results of the new multi-phase models will not change much because the population of sublimable JCH$_4$ hardly changes as discussed before. However, these abundances in the new multi-phase model 3 do deviate from that in the other new multi-phase models. At 30 K, the new multi-phase model 3 predicts slightly lower CH$_3$CCH abundance and slightly higher C$_4$H and C$_4$H$_2$ abundances than the other new multi-phase models do. Since the abundance of gas-phase CH$_4$ in the new multi-phase model 3 is lower than that in the other new multi-phase models at around 30 K, less CH$_3$CCH are formed in the new multi-phase model 3. On the other hand, we can see that C$_4$H and C$_4$H$_2$ abundances in the new multi-phase model 3 are slightly higher than those in the new multi-phase models 0-2. The discrepancy can be explained as follows. Interstitial O$_2$ and O$_3$ molecules can enter active layers upon warm-up and are hydrogenated to form OH and H$_2$O, which may enter gas phase via reactive desorption. Both gas-phase OH and H$_2$O can produce H$_3$O$^+$ ions while the electron dissociative recombination of H$_3$O$^+$ is the major destruction pathway for electrons in our models. Because the interstitial O$_2$ and O$_3$ abundances at 10 K in the new multi-phase model 3 are lower, the abundances of gas-phase OH and H$_2$O are lower at around 30 K in this model. Therefore, the new multi-phase model 3 predicts lower H$_3$O$^+$ abundance and thus higher electron abundance at around 30 K. Moreover, electron dissociative recombination is important to produce C$_4$H and C$_4$H$_2$. For instance, C$_4$H and C$_4$H$_2$ can be produced by reactions C$_4$H$_2^{+}$ + e$^{-}$ $\rightarrow$ C$_4$H + H and C$_4$H$_3^{+}$ + e$^{-}$ $\rightarrow$ C$_4$H$_2$ + H, respectively. Thus, more C$_4$H and C$_4$H$_2$ are formed in the new multi-phase model 3 than those in the other new multi-phase models at around 30 K. 
                
        \par{}Overall, the carbon-chain species abundances predicted by the basic and the new multi-phase models are typically lower than that by the two-phase model at around 30 K when WCCC occurs because gas-phase CH$_4$ in the two-phase model is much higher than that in the other models. The abundance of CH$_3$CCH in the two-phase model at around 30 K is about two orders of magnitude higher than those in the basic and the new multi-phase models, while C$_4$H$_2$ and HC$_5$N produced in the basic and the new multi-phase models at around 30 K are about one order of magnitude less than those in the two-phase model. The abundances of C$_4$H at 30 K do not vary much in different models and its abundance in the two-phase model is the highest. On the other hand, because O$_2$ or O$_3$ molecules that are buried in the bulk of ice cannot enter the active layers to be hydrogenated and form OH or H$_2$O upon warm-up in the basic multi-phase model, this model predicts lower H$_3$O$^+$ abundance and higher electron abundance than the new multi-phase models do. Therefore, although the basic new multi-phase model predicts the lowest gaseous CH$_4$ abundance at around 30 K, the carbon-chain species abundances predicted by the basic multi-phase model are not always the lowest. The abundances of C$_4$H, C$_4$H$_2$, and HC$_5$N in the basic multi-phase model are approximately the same or even slightly higher than those in the new multi-phase models 0-2.

  \section{Comparison with observations} \label{section4}
  
   \par{}In order to compare the simulation results with the observations, we adopted the `mean confidence level' method as a quantitative standard \citep{2007A&A...467.1103G,2008ApJ...681.1385H}. The confidence level $\kappa_j$ represents the agreement between the abundance in the simulation $X_j$ and the observation $X_{\mathrm{obs,}j}$ of species $j$, and $\kappa_j$ is calculated by
        \begin{align}
        \kappa_j=\mathrm{erfc} \left( \frac{|\lg X_j - \lg X_{\mathrm{obs,}j}|}{\sqrt{2}\sigma} \right),  \label{equation3}
        \end{align}
        where erfc is the complementary error function, $\sigma=1$ is the standard error. The value of $\kappa_j$ is between 0 and 1. The agreement with the observations increases as $\kappa_j$ approaches 1. We set $\sigma=1$ because we assumed the uncertainty of the observations as one order of magnitude. Therefore, $\kappa_j=0.317$ represents that the abundance of a species in a model is higher or lower by one order of magnitude than the observation. The average $\kappa$ for all the observed species towards one astronomical source can be used to quantify the global agreement between the model results and the observations. Another parameter to quantify the global agreement is the number of fits, which is the number of species that fit the observations within one order of magnitude when the value of the average $\kappa$ is the maximum. The time when the value of the average $\kappa$ is the maximum, $\kappa_\mathrm{max}$, is the optimum time $t_\mathrm{cold}$ + $\Delta t_{\mathrm{opt}}$. We compared the model results with the observations and find out $\kappa_\mathrm{max}$, $t_\mathrm{cold}$ + $\Delta t_{\mathrm{opt}}$, the number of fits, the temperature and the abundances of observed species at the time $t_\mathrm{cold}$ + $\Delta t_{\mathrm{opt}}$.

        %\begin{center}
        \begin{table*}
        %\begin{sidewaystable*}
        \caption{Observed and simulated fractional abundances of L1527. }
        \label{table4}
        \centering
          %\scalebox{0.95}{
          \begin{tabular}{lcccccccc}\\
          \hline
          \hline

          Species&L1527&RE&Two-phase&Basic multi-phase&New 0&New 1&New 2&New 3\\
            \hline
            C$_2$H&1.2(-08)\tablefootmark{a,b}&8.0(-09)&6.0(-09)&2.9(-09)&1.6(-09)&1.7(-09)&1.8(-09)&3.0(-09)\\
            C$_3$H\tablefootmark{i}&5.5(-11)\tablefootmark{c}&2.8(-10)&3.0(-10)&$\textbf{8.2(-10)}$\tablefootmark{h}&5.3(-10)&$\textbf{5.7(-10)}$&$\textbf{6.2(-10)}$&$\textbf{1.0(-09)}$\\
            C$_4$H&3.2(-09)\tablefootmark{d}&6.6(-10)&5.0(-10)&5.0(-10)&3.6(-10)&3.4(-10)&4.0(-10)&7.6(-10)\\
            C$_5$H&1.5(-11)\tablefootmark{d}&$\textbf{1.6(-10)}$&1.2(-10)&6.6(-11)&4.8(-11)&4.8(-11)&6.5(-11)&1.2(-10)\\
            C$_6$H&1.0(-11)\tablefootmark{d}&3.8(-11)&2.5(-11)&1.1(-11)&8.8(-12)&1.1(-11)&1.2(-11)&2.9(-11)\\
            C$_3$H$_2$\tablefootmark{j}&2.4(-10)\tablefootmark{b,c}&4.1(-10)&4.1(-10)&7.9(-10)&5.8(-10)&5.8(-10)&6.3(-10)&9.7(-10)\\
            C$_4$H$_2$&2.7(-11)\tablefootmark{b}&$\textbf{3.0(-09)}$&$\textbf{2.1(-09)}$&$\textbf{6.9(-10)}$&$\textbf{6.1(-10)}$&$\textbf{6.2(-10)}$&$\textbf{6.8(-10)}$&$\textbf{1.1(-09)}$\\
            C$_6$H$_2$&2.5(-12)\tablefootmark{d}&$\textbf{3.6(-10)}$&$\textbf{1.9(-10)}$&$\textbf{3.3(-11)}$&$\textbf{3.9(-11)}$&$\textbf{3.6(-11)}$&$\textbf{3.8(-11)}$&$\textbf{7.2(-11)}$\\
            CH$_3$CCH&1.0(-09)\tablefootmark{d}&3.2(-09)&1.9(-09)&$\textbf{1.3(-11)}$&$\textbf{5.8(-11)}$&$\textbf{5.5(-11)}$&$\textbf{6.5(-11)}$&$\textbf{5.3(-11)}$\\
            HC$_3$N&4.5(-10)\tablefootmark{c}&1.5(-09)&1.2(-09)&6.0(-10)&3.4(-10)&3.3(-10)&4.0(-10)&6.4(-10)\\
            HC$_5$N&1.1(-10)\tablefootmark{c}&1.7(-10)&1.1(-10)&3.1(-11)&1.9(-11)&2.6(-11)&2.8(-11)&5.1(-11)\\
            HC$_7$N&2.5(-11)\tablefootmark{d}&2.1(-11)&1.1(-11)&4.3(-12)&3.0(-12)&3.0(-12)&5.0(-12)&9.0(-12)\\
            HC$_9$N&2.3(-12)\tablefootmark{d}&2.2(-12)&1.5(-12)&2.5(-13)&2.5(-13)&5.0(-13)&5.0(-13)&1.0(-12)\\
            CCS&8.5(-11)\tablefootmark{d}&3.5(-11)&2.0(-11)&2.6(-11)&1.4(-11)&9.8(-12)&1.2(-11)&2.0(-11)\\
            HCO$_2^+$&1.0(-12)\tablefootmark{b}&6.4(-13)&1.3(-12)&$\textbf{0}\tablefootmark{k}$&1.0(-12)&1.0(-12)&1.0(-12)&5.0(-13)\\
            NH$_3$&8.3(-09)\tablefootmark{e}&8.3(-09)&1.5(-08)&1.1(-08)&1.1(-08)&1.0(-08)&9.5(-09)&8.8(-09)\\
            CO&3.9(-05)\tablefootmark{f}&2.9(-05)&2.8(-05)&$\textbf{4.4(-09)}$&$\textbf{4.5(-07)}$&$\textbf{4.0(-07)}$&$\textbf{4.7(-07)}$&$\textbf{2.3(-07)}$\\
            CN&8.0(-11)\tablefootmark{f}&$\textbf{2.6(-09)}$&$\textbf{2.7(-09)}$&$\textbf{9.8(-09)}$&$\textbf{4.3(-09)}$&$\textbf{4.4(-09)}$&$\textbf{4.5(-09)}$&$\textbf{7.0(-09)}$\\
            CS&3.3(-10)\tablefootmark{f}&1.5(-10)&9.0(-11)&1.9(-10)&2.5(-10)&2.2(-10)&2.3(-10)&2.0(-10)\\
            HCN&1.2(-09)\tablefootmark{f}&2.7(-09)&2.9(-09)&5.9(-09)&3.2(-09)&3.2(-09)&3.3(-09)&4.2(-09)\\
            HNC&3.2(-10)\tablefootmark{f}&1.9(-09)&2.1(-09)&$\textbf{5.4(-09)}$&2.5(-09)&2.6(-09)&2.6(-09)&$\textbf{3.7(-09)}$\\
            HCO$^+$&6.0(-10)\tablefootmark{f}&1.6(-09)&1.5(-09)&$\textbf{6.5(-12)}$&3.0(-10)&2.7(-10)&3.0(-10)&1.8(-10)\\
            N$_2$H$^+$&2.5(-10)\tablefootmark{f}&$\textbf{9.8(-12)}$&3.6(-11)&3.0(-10)&3.7(-10)&3.3(-10)&3.0(-10)&2.4(-10)\\
            SO&1.4(-10)\tablefootmark{f}&6.6(-11)&1.7(-10)&$\textbf{1.5(-12)}$&1.1(-10)&8.3(-11)&7.2(-11)&2.2(-11)\\
            CH$_4$&2.2-7.6(-06)\tablefootmark{g}&1.2(-05)&5.2(-06)&$\textbf{8.0(-08)}$&2.9(-07)&2.8(-07)&3.2(-07)&2.5(-07)\\  
            CH$_3$OH&1.1(-09)\tablefootmark{c}&3.5(-10)&2.7(-09)&$\textbf{2.0(-11)}$&2.2(-10)&2.1(-10)&2.3(-10)&2.1(-10)\\
          \hline
          $\kappa_{\mathrm{max}}$&-&0.604&0.616&0.398&0.531&0.532&0.542&0.506\\
          $t_{cold}+\Delta t$ ($10^5$ yr)&-&3.587&3.604&3.674&3.681&3.671&3.671&3.681\\
          $T$ (K)&-&26.37&27.33&31.58&32.03&31.39&31.39&32.03\\
          Number of fits&26&21&23&14&21&20&20&19\\
          \hline
          \end{tabular} \\
          %\end{tabular} }\\
          \tablefoot{
            \tablefoottext{a}{$a(b)=a\times10^b .$}
            \tablefoottext{b}{\citet{2009ApJ...697..769S}. }
            \tablefoottext{c}{\citet{2009ApJ...702.1025S}. }
            \tablefoottext{d}{\citet{2008ApJ...672..371S}. }
            \tablefoottext{e}{\citet{2009ApJ...699..585H}. }
            \tablefoottext{f}{\citet{2004A&A...416..603J}. }
            \tablefoottext{g}{\citet{2012ApJ...758L...4S}. }
            %\tablefoottext{h}{Italic type indicates overproduction by one order of magnitude or more. }
            \tablefoottext{h}{Bold type indicates overproduction or underproduction by one order of magnitude or more. }
            %\tablefoottext{i}{Bold type indicates underproduction by one order of magnitude or more. }
            \tablefoottext{i}{l-C$_3$H and c-C$_3$H are isomers but they are not distinguished in the reaction network. The results represent the sum of the abundances of l-C$_3$H and c-C$_3$H. So far only l-C$_3$H is observed towards L1527. }
            \tablefoottext{j}{l-C$_3$H$_2$ and c-C$_3$H$_2$ are isomers but they are not distinguished in the reaction network. The results represent the sum of the abundances of l-C$_3$H$_2$ and c-C$_3$H$_2$. The observation towards L1527 is the total abundance of l-C$_3$H$_2$ and c-C$_3$H$_2$. }
             \tablefoottext{k}{This number is below 10$^{-13}$ and for convenience we designate it as 0. }
          }       
       %\end{sidewaystable*}
       \end{table*}
       %\clearpage
       %\end{center}

   \par{}Table \ref{table4} shows the results for the comparison of the model results and the observations towards L1527. Overall, the two-phase model results fit the observations towards L1527 the best while the basic multi-phase model works the worst. Comparing the results from the two-phase model simulated with the macroscopic MC approach with the observations, the peak value of $\kappa$ is 0.616 when $t_{\mathrm{cold}}+\Delta t=3.604\times10^5\, \mathrm{yr}$ and $T=27\, \mathrm{K}$, and 23 of 26 species can fit the observations within one order of magnitude. However, $\kappa_\mathrm{max}$ for the basic multi-phase model only reaches 0.398 when $T=32\, \mathrm{K}$ while the number of fits is just 14. The results from the new multi-phase models 0, 1, and 2 agree with the observations almost equally well but the new multi-phase model 3 performs slightly worse than the other new multi-phase models. For the new multi-phase models 0, 1, and 2, $\kappa_\mathrm{max}$ is around 0.530 when $T\sim 31-32\, \mathrm{K}$ while the number of fits is 21 or 20. However, $\kappa_\mathrm{max}$ for the new multi-phase model 3 drops to about 0.510 and the number of fits is 19. In terms of fitting observations, the difference in the two-phase model results between the macroscopic MC method and the RE approach is not significant, because $\kappa_\mathrm{max}$ and the number of fits are similar. 
   
   \par{}Table \ref{table5} and \ref{table6} show the results for the comparison of the model results and the observations towards B228 and L483, respectively. The two-phase model still outperforms the basic and the new multi-phase models while $\kappa_\mathrm{max}$ for the basic multi-phase model is still the lowest among all models. The two-phase model simulated with the RE approach fits the observations towards L483 slightly better than that simulated with the macroscopic MC method, but fits the observations towards B228 slightly less well.

  %\begin{center}
        \begin{table*}
        %\begin{sidewaystable*}
        \caption{Observed and simulated fractional abundances of B228. }
        \label{table5}
        \centering
          %\scalebox{0.95}{
          \begin{tabular}{lcccccccc}\\
          \hline
          \hline

          Species&B228&RE&Two-phase&Basic multi-phase&New 0&New 1&New 2&New 3\\
            \hline
            C$_2$H&5.3(-09)\tablefootmark{a,b}&8.0(-09)&6.0(-09)&2.8(-09)&1.6(-09)&1.7(-09)&1.9(-09)&3.0(-09)\\
            C$_4$H&2.3(-09)\tablefootmark{b}&6.6(-10)&4.7(-10)&5.6(-10)&3.6(-10)&3.7(-10)&4.6(-10)&7.1(-10)\\
            C$_3$H$_2$\tablefootmark{d}&5.5(-12)\tablefootmark{b}&$\textbf{4.1(-10)}$\tablefootmark{c}&$\textbf{3.8(-10)}$&$\textbf{8.3(-10)}$&$\textbf{5.8(-10)}$&$\textbf{5.7(-10)}$&$\textbf{6.8(-10)}$&$\textbf{9.2(-10)}$\\
            C$_4$H$_2$&6.3(-12)\tablefootmark{b}&$\textbf{3.0(-09)}$&$\textbf{1.9(-09)}$&$\textbf{7.6(-10)}$&$\textbf{6.3(-10)}$&$\textbf{6.6(-10)}$&$\textbf{7.5(-10)}$&$\textbf{1.1(-09)}$\\
            CH$_3$CCH&1.7(-09)\tablefootmark{b}&3.2(-09)&1.7(-09)&$\textbf{1.2(-11)}$&$\textbf{5.7(-11)}$&$\textbf{5.2(-11)}$&$\textbf{5.9(-11)}$&$\textbf{5.2(-11)}$\\
            HC$_5$N&3.3(-11)\tablefootmark{b}&1.7(-10)&9.7(-11)&3.6(-11)&1.8(-11)&2.3(-11)&2.8(-11)&4.7(-11)\\
            HCO$_2^+$&4.5(-12)\tablefootmark{b}&6.4(-13)&3.8(-12)&$\textbf{0}$\tablefootmark{e}&2.8(-12)&2.0(-12)&2.3(-12)&7.5(-13)\\
          \hline
          $\kappa_{\mathrm{max}}$&-&0.455&0.585&0.345&0.412&0.407&0.436&0.415\\
          $t_{cold}+\Delta t$ ($10^5$ yr)&-&3.587&3.601&3.682&3.683&3.680&3.686&3.672\\
          $T$ (K)&-&26.37&27.16&32.09&32.16&31.96&32.35&31.45\\
          Number of fits&7&5&5&3&4&4&4&4\\
          \hline
          \end{tabular} \\
          %\end{tabular} }\\
          \tablefoot{
          \tablefoottext{a}{$a(b)=a\times10^b .$}
          \tablefoottext{b}{\citet{2009ApJ...697..769S}. }
          %\tablefoottext{c}{Italic type indicates overproduction by one order of magnitude or more. }
          \tablefoottext{c}{Bold type indicates overproduction or underproduction by one order of magnitude or more. }
          %\tablefoottext{d}{Bold type indicates underproduction by one order of magnitude or more. }
          \tablefoottext{d}{l-C$_3$H$_2$ and c-C$_3$H$_2$ are isomers but they are not distinguished in the reaction network. The results represent the sum of the abundances of l-C$_3$H$_2$ and c-C$_3$H$_2$. So far only l-C$_3$H$_2$ is observed towards B228. }
          \tablefoottext{e}{This number is below 10$^{-13}$ and for convenience we designate it as 0. }
          }   
       %\end{sidewaystable*}
       \end{table*}
       %\clearpage
       %\end{center}

        %\begin{center}
        \begin{table*}
        %\begin{sidewaystable*}
        \caption{Observed and simulated fractional abundances of L483. }
        \label{table6}
        \centering
          %\scalebox{0.95}{
          \begin{tabular}{lcccccccc}\\
          \hline
          \hline
          
          Species&L483&RE&Two-phase&Basic multi-phase&New 0&New 1&New 2&New 3\\
            \hline
            C$_4$H&7.0(-10)\tablefootmark{a,b}&1.5(-09)&6.9(-10)&6.5(-10)&3.5(-10)&3.9(-10)&4.5(-10)&6.5(-10)\\
            C$_3$H$_2$\tablefootmark{j}&1.8(-10)\tablefootmark{c}&7.7(-10)&4.8(-10)&8.7(-10)&5.5(-10)&5.7(-10)&6.6(-10)&8.8(-10)\\
            HC$_3$N&5.2(-10)\tablefootmark{d}&1.5(-09)&1.4(-09)&5.3(-10)&3.4(-10)&3.3(-10)&3.8(-10)&5.6(-10)\\
            HC$_5$N&2.9(-10)\tablefootmark{d}&2.5(-10)&1.5(-10)&3.9(-11)&$\textbf{2.3(-11)}$\tablefootmark{i}&$\textbf{1.8(-11)}$&$\textbf{2.5(-11)}$&4.4(-11)\\
            CCS&2.1(-10)\tablefootmark{d}&2.9(-11)&2.1(-11)&3.0(-11)&$\textbf{1.4(-11)}$&$\textbf{1.3(-11)}$&$\textbf{1.2(-11)}$&$\textbf{1.9(-11)}$\\
            CO&1.4(-05)\tablefootmark{e}&2.9(-05)&2.8(-05)&$\textbf{4.7(-09)}$&$\textbf{4.5(-07)}$&$\textbf{4.3(-07)}$&$\textbf{4.9(-07)}$&$\textbf{2.3(-07)}$\\
            HCN&2.0(-09)\tablefootmark{e}&2.3(-09)&2.9(-09)&5.5(-09)&3.3(-09)&3.0(-09)&3.2(-09)&4.1(-09)\\
            HNC&3.9(-10)\tablefootmark{e}&1.6(-09)&2.0(-09)&$\textbf{5.0(-09)}$&2.6(-09)&2.5(-09)&2.6(-09)&3.6(-09)\\
            CS&6.8(-10)\tablefootmark{e}&1.3(-10)&9.1(-11)&1.9(-10)&2.6(-10)&2.4(-10)&2.2(-10)&1.9(-10)\\
            SO&2.9(-10)\tablefootmark{e}&7.6(-11)&1.7(-10)&$\textbf{1.0(-12)}$&1.0(-10)&8.5(-11)&7.1(-11)&$\textbf{2.5(-11)}$\\
            HCO$^+$&2.0(-09)\tablefootmark{e}&1.6(-09)&1.5(-09)&$\textbf{5.0(-12)}$&3.0(-10)&3.0(-10)&3.2(-10)&$\textbf{1.7(-10)}$\\
            N$_2$H$^+$&7.5(-10)\tablefootmark{e}&$\textbf{1.0(-11)}$&$\textbf{3.0(-11)}$&3.1(-10)&3.7(-10)&3.6(-10)&3.2(-10)&2.4(-10)\\
            CN&8.3(-10)\tablefootmark{e}&1.0(-09)&2.4(-09)&$\textbf{9.2(-09)}$&4.4(-09)&4.4(-09)&4.4(-09)&6.8(-09)\\
            CH$_3$OH&4.3(-10)\tablefootmark{f}&2.1(-10)&3.0(-09)&$\textbf{1.8(-11)}$&2.2(-10)&2.0(-10)&2.2(-10)&2.1(-10)\\
            NH$_3$&2.5(-08)\tablefootmark{d}&4.9(-09)&1.4(-08)&1.1(-08)&1.1(-08)&1.0(-08)&9.5(-09)&9.0(-09)\\
            H$_2$CO&8.3(-10)\tablefootmark{g}&6.8(-09)&5.1(-09)&$\textbf{3.6(-11)}$&7.0(-09)&6.0(-09)&6.6(-09)&2.5(-09)\\
            HCCO&7.2(-12)\tablefootmark{h}&$\textbf{1.1(-13)}$&$\textbf{0}$\tablefootmark{k}&$\textbf{0}$\tablefootmark{k}&3.8(-11)&2.4(-11)&1.5(-11)&1.0(-12)\\
            CH$_3$CHO&5.3(-11)\tablefootmark{h}&1.5(-10)&$\textbf{3.8(-12)}$&8.3(-12)&$\textbf{1.3(-12)}$&$\textbf{7.5(-13)}$&$\textbf{5.0(-13)}$&$\textbf{0}$\tablefootmark{k}\\
            HCO&4.5(-11)\tablefootmark{h}&1.0(-10)&3.6(-10)&$\textbf{5.0(-13)}$&3.5(-11)&3.0(-11)&2.7(-11)&1.3(-11)\\
          \hline
          $\kappa_{\mathrm{max}}$&-&0.605&0.563&0.383&0.549&0.544&0.551&0.488\\
          $t_{cold}+\Delta t$ ($10^5$ yr)&-&3.652&3.617&3.700&3.676&3.695&3.672&3.684\\
          $T$ (K)&-&30.19&28.08&33.28&31.71&32.94&31.45&32.22\\
          Number of fits&19&17&16&10&15&15&15&14\\
          \hline
          \end{tabular} \\
          %\end{tabular} }\\
          \tablefoot{
          \tablefoottext{a}{$a(b)=a\times10^b .$}
          \tablefoottext{b}{\citet{2009ApJ...697..769S}. }
          \tablefoottext{c}{\citet{1998ApJ...506..743B}. }
          \tablefoottext{d}{\citet{2009ApJ...699..585H}. }
          \tablefoottext{e}{\citet{2004A&A...416..603J}. }
          \tablefoottext{f}{\citet{2002A&A...389..908J}. }
          \tablefoottext{g}{\citet{2000A&A...359..967T}. }
          \tablefoottext{h}{\citet{2015A&A...577L...5A}. }
          %\tablefoottext{i}{Italic type indicates overproduction by one order of magnitude or more. }
          \tablefoottext{i}{Bold type indicates overproduction or underproduction by one order of magnitude or more. }
          %\tablefoottext{j}{Bold type indicates underproduction by one order of magnitude or more. }
          \tablefoottext{j}{l-C$_3$H$_2$ and c-C$_3$H$_2$ are isomers but they are not distinguished in the reaction network. The results represent the sum of the abundances of l-C$_3$H$_2$ and c-C$_3$H$_2$. So far only l-C$_3$H$_2$ is observed towards L483. }
          \tablefoottext{k}{These numbers are below 10$^{-13}$ and for convenience we designate them as 0. }
          }  
       %\end{sidewaystable*}
       \end{table*}
       %\clearpage
       %\end{center}

   \par{}Gaseous CH$_4$ abundances predicted by the basic and the new multi-phase models at $t_{\mathrm{cold}}+\Delta t_{\mathrm{opt}}$ are around one order of magnitude lower than the observations towards L1527, while the two-phase model can reproduce the observed CH$_4$ abundance pretty well. Because gaseous CH$_4$ is the key molecule for the synthesis of carbon-chain species based on the WCCC mechanism, the abundances of carbon-chain species predicted by the basic and the new multi-phase models are typically lower than that by the two-phase model. However, the agreement between the model results and the observations varies depending on the carbon-chain species. The abundances of CH$_3$CCH predicted by the basic multi-phase model are about two orders of magnitude lower than the observations towards L1527 and B228. The abundances of CH$_3$CCH predicted by the new multi-phase models are about a factor of four larger than that in the basic multi-phase model, but they are still more than one order of magnitude lower than the observations. On the other hand, the abundances of C$_2$H and HC$_3$N predicted by the basic and the new multi-phase models agree reasonably well with the observations towards all known sources. The two-phase model overproduces C$_4$H$_2$ and C$_6$H$_2$ by about two orders of magnitude. Although the abundances of these two species drop in the basic and the new multi-phase models, they are still more than one order of magnitude higher than the observations towards L1527 and B228. In addition to CH$_4$ and carbon-chain species, another species that is severely underproduced by the basic and the new multi-phase models is CO. The observed CO abundances towards L1527 and L483 are four orders of magnitude higher than that predicted by the basic multi-phase model. The abundances of CO increase by two orders of magnitude in the new multi-phase models, but they are still two orders of magnitude lower than the observations.

  \section{Discussions and conclusions} \label{section5}
  \par{}We investigated WCCC using the two-phase model, the basic and the new multi-phase models in this work. Because the macroscopic MC method was adopted in the simulations, we also paid attention to the significance of the finite size effect in the simulations. The results of the basic and the new multi-phase models are compared with that of the two-phase model. We find that gaseous CH$_4$ abundance predicted by the basic multi-phase model is more than one order of magnitude lower than that predicted by the two-phase model when the temperature is around 30 K because most CH$_4$ is frozen on grains. In addition to CH$_4$ molecules in the active layers, interstitial CH$_4$ can also sublime at 30 K in the new multi-phase models, thus the abundances of gaseous CH$_4$ in the new multiphase models are around a factor of three larger than that in the basic multi-phase model. However, gaseous CH$_4$ abundances in the new multi-phase models are still around one order of magnitude lower than that in the two-phase model. Because CH$_4$ is important to synthesize carbon-chain species in the gas phase, the abundances of carbon-chain species in the two-phase model are the highest among all models. On the other hand, the increase of H$_3$O$^+$ abundance can decrease the abundance of electrons. Meanwhile, electrons participate in the reactions of producing carbon-chain species. Therefore, carbon-chain species abundances in the basic multi-phase model at 30 K may be even higher than those in the new multi-phase models, because the H$_3$O$^+$ abundance in the basic multi-phase model is the lowest among all models. In addition, the significance of the finite size effect depends on the duration of the cold stage. If $t_{\mathrm{cold}} >3\times 10^5$ yr, the finite size effect is not important. However, if $t_{\mathrm{cold}} < 3\times 10^5$ yr, the difference in the results of the two-phase model between the RE approach and the macroscopic MC method is large. 

  \par{}Although the basic multi-phase model is more realistic than the two-phase model because icy species are distinguished, the abundances of species predicted by the basic multi-phase model agree worse with the observations than that by the two-phase model. The new multi-phase models do improve the agreement with the observations. However, a few species such as CH$_4$ and CH$_3$CCH are still severely underproduced by the new multi-phase models. We discuss the reasons for this below. 
  
  \par{}First, the UV radiation field in the cold stage may be stronger than that used in our simulations so that more interstitial species are formed, thus CH$_4$ abundance may increase upon warm-up. Second, physical conditions can affect the chemical model results. For instance, we ran test simulations with the physical model used by \citet{2008ApJ...674..984A}. We find that the abundances of gaseous CH$_4$ and CO agree well with the observations towards L1527. However, overall the abundances of species predicted by all chemical models using this physical model agree worse with the observations. The reason may be that the pre-collapse duration in this physical model, $10^6$ yr, may be too long. Finally, more rigorous lattice microscopic MC simulations show that abundant holes may exist in the ice mantle \citep{2013ApJ...778..158G}. If there are abundant holes in the ice mantles on grain surfaces, more CH$_4$ molecules in the bulk of ice might diffuse into the active layers upon warm-up, thus more gaseous carbon-chain species may be produced. Therefore, WCCC may be used to test the validity of surface chemical models. 

  \par{}It is found that the population of species that can diffuse in the bulk of ice controls the efficiency of the synthesis of terrestrial COMs in the ice mantle \citep{2018arXiv180902419L}. On the other hand, the population of solid CH$_4$ that are mobile in the bulk of ice is also crucial for WCCC. Therefore, more studies on WCCC will also help to discover the formation of terrestrial COMs in various astronomical sources. \\

  \begin{acknowledgements}
We thank our referee's constructive comments to improve the quality of the manuscript. The authors are grateful to Donghui Quan and Xia Zhang for the rate equation simulations. We also thank Eric Herbst, Ling Liao, Yang Lu, Longfei Chen, and Fujun Du for helpful discussions and valuable comments. We thank George E. Hassel for providing chemical reaction networks he used for WCCC simulations. We also thank Yuri Aikawa for providing the physical model she used for WCCC simulations. This work is supported by the National Key R\&D Program of China (NO. 2017YFA0402701). Q.C. acknowledges the support from the National Natural Science Foundation of China (NSFC) grant No. 11673054.
  \end{acknowledgements}
  
  %\bibpunct{(}{)}{;}{a}{}{,} % to follow the A&A style
  % for the bibliography, at the end
  \bibliographystyle{aa} % style aa.bst
  %\bibliography{references1} % your references Yourfile.bib
  \bibliography{references5}

  %\begin{appendix}
  %\section{Title of the first appendix}
  %\section{Title of the second}
  %\end{appendix}
                
\end{document}